\documentclass[12pt,preprint]{aastex}

\newcommand{\Lya}{Lyman-$\alpha$}

\newcommand{\chisq}{$\chi^{2}$}
\newcommand{\chisqnu}{$\chi^{2}_{\nu}$}
\newcommand{\kms}{km s$^{-1}$}
\newcommand{\kmsmpc}{km s$^{-1}$ Mpc$^{-1}$}
\newcommand{\hfifty}{$h_{50}^{-1}$}

\newcommand{\gamres}{$\Gamma_{res}$}

\newcounter{n}

\slugcomment{Submitted to the Astrophysical Journal}

\shorttitle{\Lya\ ``Forest'' Simulations}
\shortauthors{Petry et al.}

\begin{document}
\doublespace

\title{Comparing Simulations and Observations of the Lyman-$\alpha$ Forest\\
I. Methodology }

\author{Cathy E. Petry and Chris D. Impey}
\affil{Steward Observatory, The University of Arizona, Tucson, AZ 85721}
\email{cpetry@as.arizona.edu, cimpey@as.arizona.edu}

\author{Neal Katz}
\affil{Department of Astronomy, University of Massachusetts, Amherst, MA 01003}
\email{nsk@kaka.astro.umass.edu}

\author{David H. Weinberg}
\affil{Department of Astronomy, The Ohio State University, Columbus, OH 43210}
\email{dhw@astronomy.ohio-state.edu}

\and

\author{Lars E. Hernquist}
\affil{Department of Astronomy, Harvard University, Cambridge, MA 02138} 
\email{lars@cfa.harvard.edu}
\email{}
\email{}

\begin{abstract}
We describe techniques for comparing spectra extracted from cosmological
simulations and observational data, using the same methodology to link \Lya\ 
properties derived from the simulations with properties derived from 
observational data.  The eventual goal is to measure the coherence or 
clustering properties of \Lya\ absorbers using observations of quasar pairs and
groups. We quantify the systematic underestimate in opacity that is inherent
in the continuum fitting process of observed spectra over a range of resolution 
and $SNR$.  We present an automated process for detecting and selecting
absorption features over the range of resolution and $SNR$ of typical 
observational data 
on the \Lya\ ``forest''.  Using these techniques, we detect coherence over 
transverse scales out to 500 \hfifty\ kpc in spectra extracted from a 
cosmological simulation at $z=2$.

\end{abstract}


\section{INTRODUCTION}

The numerous lines of Lyman-$\alpha$ absorption that appear in the spectra
of quasars are proving to be excellent cosmological probes. Hydrodynamic
simulations of the universe reveal an evolving network of sheets, filaments
and halos caused by complex gravitational dynamics in the expanding universe.
The Lyman-$\alpha$ ``forest'' only identifies the neutral component of the
intergalactic medium, but the absorbers are well understood tracers of the 
overall mass distribution \citep{her96, mir96, rau98}. The simulations show 
that the physical state of the diffuse gas causing Lyman-$\alpha$ absorption 
is relatively simple, enabling the use of analytic models to relate the 
absorption to the underlying mass and velocity fields \citep{mcg90,
bi93, gne96}. The combination of theory and simulation has allowed the
Lyman-$\alpha$ forest to be used for measurements of the baryon fraction
\citep{rau97,wei97}, the mass density \citep{wei99}, the amplitude of mass
fluctuations \citep{gne98,nus00}, the mass power spectrum \citep{crof98,cro99,
cro00,mcd00}, the thermal history of the IGM \citep{hae98,mcd00,ric00,sch00}, 
the chemical evolution of the Universe \citep{co99, aa00a, aa00b}, and the
metallicity of the IGM \citep{rhs97, hell97,dav98}.

On the observational side, the detection of coincident absorption lines in 
the spectra of quasar pairs provided the first evidence that \Lya\ absorbers 
have a large transverse extent \citep{din94, bec94, din98, cro98}.  In 
principle, this idea can be extended with the use of quasar groups to thread 
a contiguous volume and provide 3D ``tomography'' of the absorbers, revealing
the variation of coherence and homogeneity with redshift. In practice, however,
quasars of the appropriate brightness, redshift, and angular separation are 
hard to find. Moreover, at $z \la 2$, the opacity drops steadily
and the Lyman-$\alpha$ forest thins out into a savannah, in agreement
with numerical simulations \citep{dave99}. Over most of the Hubble time,
the absorbers can only be observed in the vacuum ultraviolet with the twenty
times smaller collecting area of the Hubble Space Telescope (HST). Studies
of the low redshift Lyman-$\alpha$ forest are still very much limited by
the quality of the available data.

\begin{figure}
\epsscale{.80}
\plotone{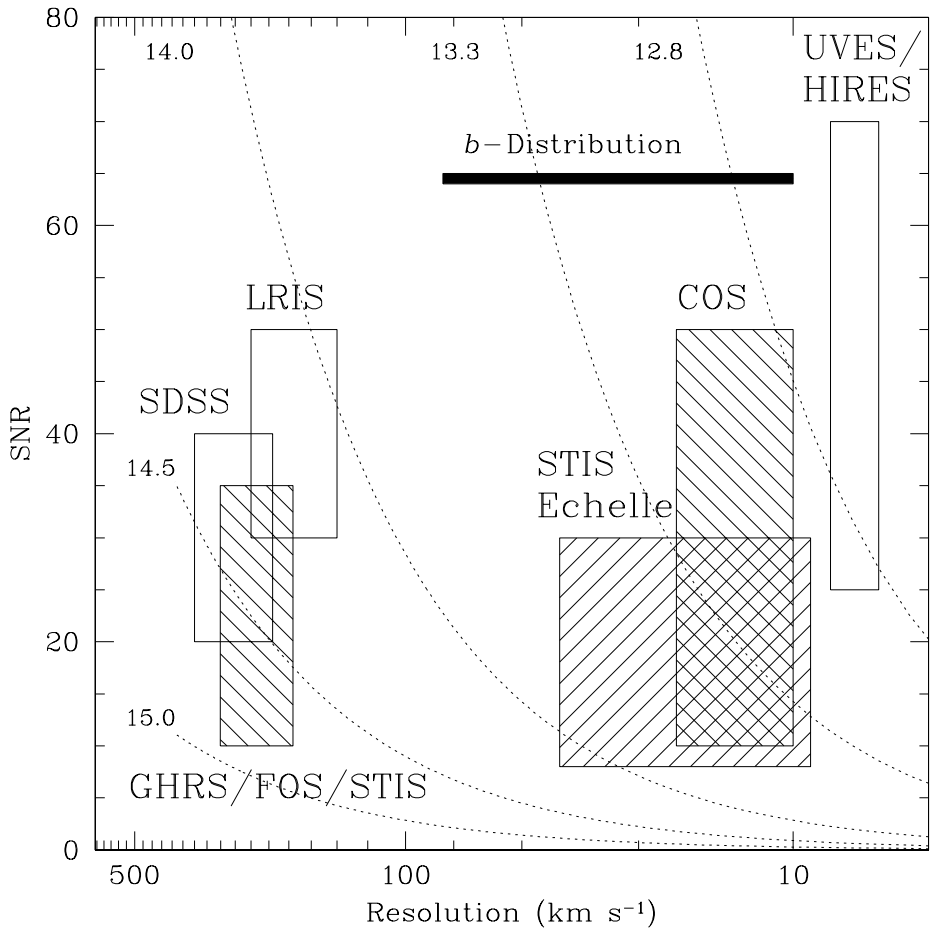}
\caption{A schematic view of the range in $SNR$ and resolution of the major 
examples of quasar spectra from pioneering observational facilities. Open 
boxes refer to ground-based observations of
Lyman-$\alpha$ with $z \gtrsim 1.6$; shaded boxes refer to the unique capabilities
of HST at lower redshifts. The horizontal bar shows the (redshift-independent) 
range of Doppler parameters, as measured with echelle data.
The dotted lines show the detection limits for a $5\sigma_{det}$ line having hydrogen
column density as labeled, $\log N=15.0,14.5,14.0,13.3,12.8$ atoms 
cm$^{-2}$. The detection limits are computed assuming the spectral 
dispersion is one-third of the instrumental resolution.
\label{fig-7}}
\end{figure}

Figure~\ref{fig-7} summarizes the state of observational capabilities for
studies of the Lyman-$\alpha$ forest. The boxes give the approximate bounds 
in resolution and $SNR$ of several different instrument and telescope 
combinations.  Open boxes show examples of ground-based facilities that can 
only measure Lyman-$\alpha$,
at $z \gtrsim 1.6$; shaded boxes show past, present and future HST
instruments. The horizontal bar shows the range of observed Doppler parameters
of the absorbing gas, which is independent of redshift \citep{pen00b}. Only
echelle observations are sensitive to the thermal and hydrodynamic 
properties of the absorbing gas. The best quality data come from Keck/HIRES
and VLT/UVES. At lower redshift, the STIS echelle mode
will be supplanted by the Cosmic Origins Spectrograph
(COS). There is a large jump in resolution down to the level (1-2 \AA)
where a larger number of targets is available and a larger redshift path
length can be surveyed. The Keck/LRIS region is typical of most moderate
resolution ground-based data. The volume of existing data will be eclipsed by 
the $\sim 10^4$ high redshift quasar spectra that will emerge from the Sloan 
Digital Sky Survey \citep{yor00}. At lower redshift, the largest single data 
set comes from the HST Quasar Absorption Line Key Project \citep{jan98,wey98}.

The goals of this paper are to (a) define a set of automated procedures 
that can be used to analyze data (with a wide range of spectral resolution, 
$SNR$, and physical separations between paired lines of sight) and spectra 
extracted from simulations, (b) form a bridge between the historically 
different methodologies of observers and simulators, and (c) facilitate 
the comparison between observations and simulations of the Lyman-$\alpha$
forest, in order to derive cosmological constraints.

The major challenge to the first goal is the fact that absorption features
in a quasar spectrum must be measured with respect to an unknown continuum,
and in the presence of broad emission features. This is usually accomplished with 
a low-order function fitted to the continuum, but no procedure is fully robust 
in dealing with 
broad absorption features or the region just blue-ward of the Lyman-$\alpha$ 
emission line. Even the comprehensive software of the HST Key Project
cannot automatically deblend complex spectral regions \citep{kp2}. By
contrast, spectra extracted from the hydrodynamic simulations yield opacities 
directly with respect to a predetermined continuum.  In addition, observers have
traditionally treated absorption in terms of discrete features modeled
by Voigt profiles, which in turn must be convolved with the instrumental 
response function. This approach is observationally successful, particularly
at low redshift where spectral features have low number density and appear 
isolated in the spectra. Regardless of whether or not a Voigt profile provides 
a suitable description of the absorbing gas, there is evidence from extracted
simulation spectra that much of the opacity is associated with shallow and
smoothly varying absorption. This is commonly referred to as the 
fluctuating Gunn-Peterson approximation \citep{her96,wkh98}.

With regard to the second goal, there are substantial differences
between the extraction of information from optical or UV spectra, and
the the extraction of information from a spectrum created as a product
of a cosmological simulation. Observational spectra have finite $SNR$ and
resolution, each one spans a large 1-dimensional redshift range, and the
Lyman-$\alpha$ lines must be identified in spectral regions contaminated
with metal lines from higher column density absorbers. Details of the method 
for culling out metal lines from the Lyman-$\alpha$ forest will be given in 
Paper II.  Spectra extracted
from hydrodynamic simulations have no noise and extremely high resolution,
although the spectral features result from gas kinematics that map onto
velocity space in a complex way. By contrast to the $\sim$ Gpc or larger 1D path
of the Lyman-$\alpha$ forest in a quasar spectrum, the simulation boxes
span 20-40 \hfifty\ Mpc and are aliased for large scale structure measurements
on scales larger than 5-10 \hfifty\ Mpc. However, the simulations are ideal 
for 3D measures
of the gas distribution since $\sim 10^5$ independent spectra can be
extracted, whereas 3D observations of absorber structure are limited
by the rarity of suitable quasar pairs.

The third goal of this project will be realized with the second and
subsequent papers in this series. In the second paper, we will compare
coherence measures of absorbers at $z \sim 2$ with extractions from
simulations. In future papers, we will extend the work to lower redshifts
and investigate the dependence of the predicted absorber properties on
cosmological parameters. This paper sets the stage by describing the
techniques for automated and consistent comparisons between observations
and simulations of the Lyman-$\alpha$ forest. In section 2, we discuss
the detection and selection of absorption features. In section 3, we 
discuss the measurement of physical properties of the absorbers. In
section 4, we discuss results from single and multiple sight-lines, 
including a comparison between line counting and measures that use
the entire information contained in the spectra. Section 5 contains 
a brief summary.

\section{ABSORPTION LINE SELECTION}

\subsection{Extracting Spectra from the Simulation}

A numerical simulation of a $z=2$ cold dark matter (CDM) universe, with
$H_0=50$ \kmsmpc\ , normalized to yield a present-day rms mass fluctuation
of $\sigma_{16}=0.7$ in spheres of radius 16 \hfifty\ Mpc, was performed by
\citet{her96} using a method based on smoothed-particle hydrodynamics, or
SPH \citep{hk89, kat96}. The simulation volume is a periodic cube of comoving
size 22.2 \hfifty\ Mpc drawn randomly from a CDM universe with $\Omega_m=1$ 
and baryon density $\Omega_b=0.05$.  The simulation includes the effects of 
a uniform photoionizing radiation field, where radiative heating and cooling
rates are computed assuming optically thin gas in ionization equilibrium with
this field. There are $64^3$ SPH particles and an equal number of dark matter
particles; the masses of the individual particles are $1.45\times10^8 M_{\sun}$
and $2.8\times10^9 M_{\sun}$, respectively. The gas resolution varies from
$\sim 5$ kpc in the highest density regions to $\sim 200$ kpc in the lowest
density regions. Although more recent simulations have been performed with
higher spatial resolution and with the parameters of the currently favored
cosmological model ($H_0=70$ \kmsmpc\ , $\Omega_m=0.35$, $\Lambda=0.65$), 
most of the issues that relate to comparisons with observations can be 
illustrated with this single simulation.

Spectra are extracted from the simulation cube by computing
the imprint of the density, temperature, and velocity of the neutral gas
fraction associated with each SPH gas particle on a flat continuum (for more
details, see Katz et al. 1996).  One
thousand sets of spectra were extracted from the simulation cube by randomly
selecting 1000 positions from the x-projection. For each 
of the 1000 primary lines of sight (PLOS), six adjacent spectra were extracted 
that are offset 33.3, 100, 233, 400, 667, and 1000 $h_{50}^{-1}$ kpc in proper
distance from each PLOS. The azimuth angle of the adjacent spectra was varied
randomly from one primary sightline to another. This strategy allows the examination 
of coherence of the absorption features in the extracted spectra over 
these (and other intermediate) transverse separations, in addition to enabling 
comparisons with
observational data for quasar pairs having a range of separations. The transverse 
separations between the lines of sight range from less than the resolution of
the simulation to a level where absorber coherence is expected to be very weak.

\begin{figure}
\epsscale{.70}
\plotone{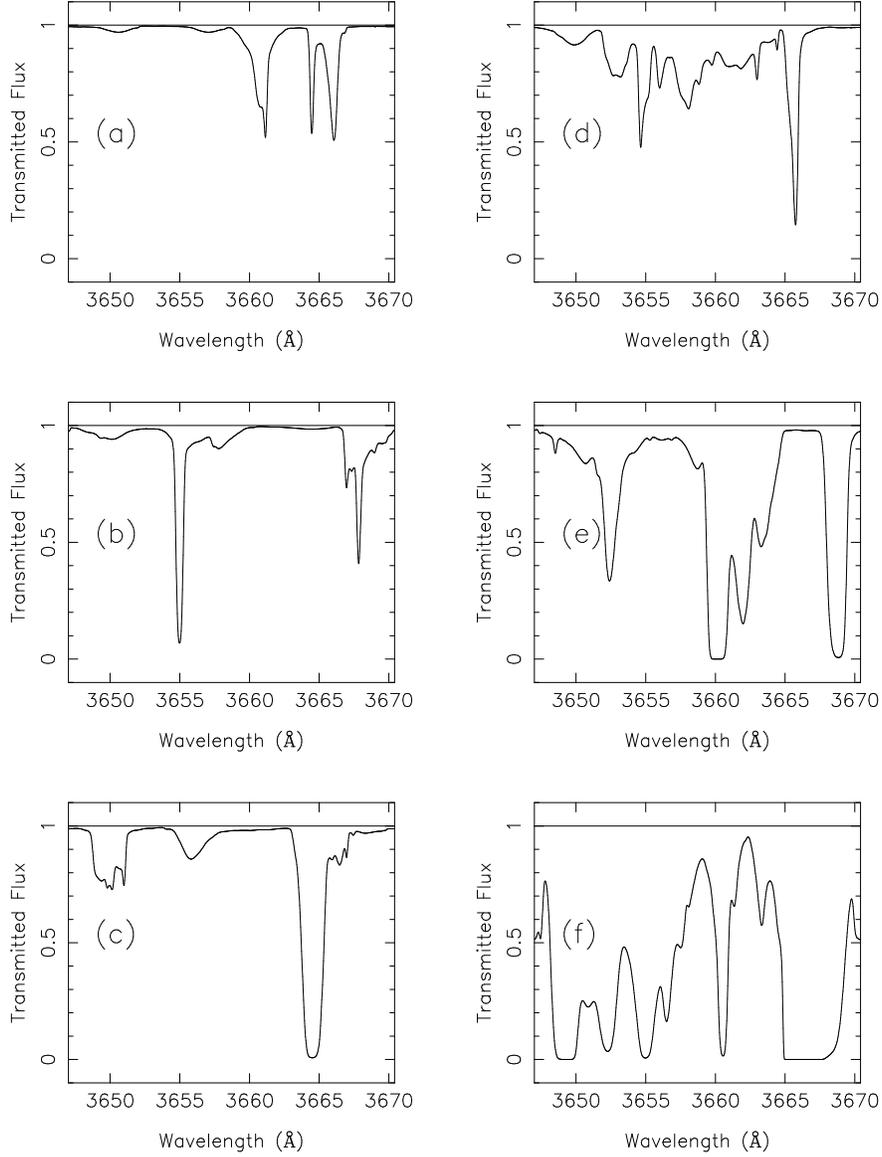}
\caption{Examples of undegraded (i.e. raw) spectra extracted from the
simulation. 
The mean transmitted flux at $z=2$ is 0.83; panels $(c)$ and $(d)$ have a
typical amount of absorption. 
The mean opacities from panels $(a)$ to $(f)$ are 0.06, 0.07, 0.13, 0.15,
0.34, 1.06.
The percent of simulated spectra with opacity less than that shown in 
panels $(a)$ through $(f)$ is 1\%, 6\%, 37\%, 47\%, 95\%, and 99\%, respectively.
\label{fig-9}}
\end{figure}

Each spectrum is sampled with 1000 data points spanning 1924.5 \kms\ , which 
corresponds to a spectral range of 23.4 \AA\ at $z=2$. Figure~\ref{fig-9}
shows six spectra extracted from the simulation illustrating the range of 
absorption features, with the mean opacity increasing going from panel $(a)$ 
to $(f)$. Panels $(c)$ and $(d)$ have a typical opacity for extracted spectra 
at this redshift, and less than 1\% of the sight-lines are as heavily absorbed 
as the example in panel $(f)$.

\subsection{Degrading the Extracted Spectra}

One objective of this paper is to compare absorber properties measured from 
the observed spectra to those measured from the spectra extracted from
the simulation. For a 
fair comparison to be made, the extracted spectra are processed to reflect 
the inevitable limitations of observational data due to instrumental effects 
of noise and resolution. The spectra are extracted from the simulation cube 
in terms of optical depth as a function of velocity, and are converted to 
transmission, $T=e^{-\tau}$, as a function of wavelength in 
Angstroms, $\lambda_i = \lambda_{Ly\alpha} (1+z) (1+v_i/c)$, where $z=2$.

As can be seen in Figure~\ref{fig-7}, the bulk of the UV data on the
Lyman-$\alpha$ forest, which contains information about the IGM over the
last three-quarters of the age of the universe, is of rather modest $SNR$ 
and resolution. This means that the complex dynamical processes that imprint
themselves on the absorber profile are not visible at low redshift, even
in the few published echelle spectra \citep{pen00a, tri00}. With non-echelle
spectra, the spectral profile is dominated by the (typically) Gaussian shape of 
the instrumental profile, and Doppler parameters cannot be measured. For 
the purposes of this paper and the following Paper II, we chose a range
of values for the resolution, \gamres, and $SNR$ typical of most published
quasar spectra. To span the parameter space of available data at any 
redshift we chose $SNR$ = 10, 30, and 100, 
and \gamres = 5, 20, 80, and 300 \kms, and form twelve 
realizations of a single spectrum extracted from the simulation, each with 
a different combination of $SNR$ and resolution.

The extracted spectra are convolved with a Gaussian profile having a FWHM, 
\gamres, with each of the four values chosen to represent the instrumental 
resolution of the available observational data. The velocity resolution of 
the extracted spectra ($\sim 2$ \kms) is much higher than the resolution of observed 
quasar spectra, the range of which is represented by the four values 
5, 20, 80 \& 300 \kms. We resample the extracted spectra to 3 per resolution 
element by spline interpolation. Poisson noise is added to each spectrum by 
scaling the intensity at each pixel to the $(SNR)^{2}$, adopting a Poisson 
deviate with that value as the mean, and then renormalizing the spectrum.  

Figure~\ref{fig-8} shows two different extracted spectra each having an 
opacity similar to the mean at $z=2$, exemplifying the typical absorption
morphologies at that redshift. The twelve realizations for each individual
extracted spectrum show the degraded appearance due to the various choices 
of $SNR$ and resolution. The highest resolution, highest $SNR$ spectrum (top 
right panel in each case) is a fairly close match to the original spectrum 
extracted from the simulation.  The best published spectra from the Keck
telescopes appear like those in the top right panels, but there are a significant
number of HST spectra that are similar in quality to the lower left panels.
These plots clearly show the separate effects
of $SNR$ and resolution on the appearance of the spectrum.

\subsection{Fitting Continua to the Extracted Spectra}

The measurement of absorption features in quasar spectra, whether by line
profile fitting or by using a type of flux statistic, must be 
performed relative to the quasar continuum. Continuum fitting is generally 
done by iteratively fitting the data points in a spectrum, eliminating 
negative or downward excursions from the fit, and refitting. The position
of the final adopted continuum depends on the $SNR$ of the spectrum and, 
more importantly, on the total flux absorption (which is directly related 
to the line density) and its distribution across the spectrum. Such a 
procedure only approaches the ``true'' continuum in the limit of infinite
signal to noise, and even then has the potential to miss weak or shallow
absorption features. Figure~\ref{fig-8} demonstrates how features that can 
be seen clearly in the highest $SNR$, highest resolution spectra are lost in 
lower $SNR$, lower resolution spectra.

Given that there are no regions of truly zero absorption, it is essentially 
{\it guaranteed} that line and continuum fitting will miss some portion of the
Lyman-$\alpha$ opacity. The systematic underestimation of the continuum
level will also tend to underestimate the equivalent width and therefore the 
column density of fitted lines too. We use extracted spectra from the 
simulation to calibrate the size of this effect, and measure it as a function 
of $SNR$ and resolution. 

With many thousands of spectra to process, an automated procedure is
required. After the simulation spectra are manipulated to mimic observational data, 
the continuum must be estimated as for the observed spectra. Any algorithm
must be simple conceptually and should be able to fit continua automatically
for spectra with a range in absorber opacity corresponding to a redshift range 
of $0<z<3$, and for spectra having a range of signal-to-noise ratios, 
$10< SNR <100$.  Any method implicitly assumes that the data retains some of 
the quasar's emission continuum, and this implies that continua fit to highly 
absorbed regions will be less representative of the true continuum. Mean 
absorption opacity and $SNR$ are the primary factors in determining how well 
the true (or input) continuum can be recovered.

Fitting a cubic spline is a standard method of estimating continua for observed 
spectra \citep{kp2}. Splines offer the advantage of being able to fit a smooth,
continuous function through a large number of data points. In quasar spectra, 
the large available wavelength range acts to define a smooth, slowly-varying 
continuum. However, the simulation volume offers only a short spectral range, and 
spline fitting to short extracted spectral segments with significant absorption
can produce steeply-sloped continua.  We suppress this effect in the case
of the extracted spectra by ``tripling'' the spectrum, or lining up three 
identical spectra, taking advantage of the fact that the simulation cube is 
periodic.  This is merely a practical mechanism for dealing with the edge 
effects in continuum and absorption line fitting --- for the analysis, only 
lines whose centers fall in the central third of the new spectrum (the
original extracted spectrum) are included.

To estimate a continuum computationally, representing what one might fit 
``by eye,'' we implement a two step process. The first step is to iteratively
fit a straight line to 23 \AA\ sections of a spectrum (corresponding to the
independent spectral segment length for the extracted spectra), providing an 
estimate of the amplitude of the continuum over that wavelength range.  
This is a reasonable way to proceed because emission continua of quasars 
can be closely approximated by series of linear segments over $\sim 20$ \AA\ 
scales, except in the vicinity of \Lya\ emission. All data points are 
weighted equally and an iterative process rejects points that deviate 
negatively by $2\sigma$ from the current fit, excluding them from 
subsequent fits.  The rejection process converges after only four cycles.

The method of continuum fitting used by the HST Absorption Line Key Project
divides a spectrum into bins and fits a spline to the average flux in 
these bins.  However, this bin size is larger than the length of our extracted
simulation spectra, so the first step described above provides our first
estimate to the continuum. The second step produces a smoothly varying 
continuum over the length of the spectrum using the amplitude of each of the 
fitted line segments from the first step.
A single point is created for each of the three segments by evaluating the 
amplitude of the fitted straight line at the average wavelength of each 
segment. A cubic spline is fit through these points for the length of the 
spectrum. Visual inspection of spectra with a range of flux decrements
demonstrates that this two step method results in a more satisfactory fit
to the continuum than the Key Project method, especially for heavily absorbed
spectra, as the first step has the effect of ``floating'' the continuum to 
what appears ``by eye'' to be a more appropriate level.

To evaluate how well the algorithm works for the observational spectra, we 
compare the final fitted continua using the new two-step process to the 
continuum fit using the method used by the HST Absorption Line Key Project. 
Some manual adjustment of the continuum is necessary in 
the vicinity of the steeply sloped Lyman-$\alpha$ emission features. 
Four of the six spectra to be analyzed in Paper II of this series --- PG~1343+264A,B, 
LB~9605, and LB~9612 --- were studied using both methods, and the ratio of the 
two fitted continua is unity at all wavelengths to well within the errors. 
This demonstrates 
that no significant systematic effects are introduced by the continuum 
fitting algorithm. We can also infer that absorption line parameters
returned by the two methods will not significantly differ.

To show how $SNR$ and resolution affect the level of the fitted continuum,
we fit continua using our method to 1000 randomly-selected
spectra extracted from the simulation.  We then compare the mean opacity 
computed for each of 1000 undegraded (i.e. raw) spectra extracted from the 
simulation to the mean opacity calculated relative to the fitted continuum for 
each degraded spectra in each of the 12 realizations of $SNR$ and \gamres.
Figure~\ref{fig-11} shows the distribution of the difference of these two
quantities for all 1000 spectra and for each realization.  This difference
is equivalent to the difference between the input continuum level (unity, by
definition) and the fitted continuum level, converted into an opacity decrement,
$\Delta \tau$.  

An opacity decrement is the unavoidable consequence of any fitting procedure where
the continuum is not known {\it a priori} --- as is the case for all 
observational data. The quantity $\Delta \tau$ therefore represents ``lost''
opacity, and it has two components.  The first is due to the
fact that strong absorbers will have their column densities slightly 
underestimated owing to the lower placement of the continuum.  The second
is due to the sum of broad and/or low level absorption that cannot be 
recovered by fitting a noisy continuum (the Gunn-Peterson approximation).
See Figure~\ref{fig-9}$d$ for a good example.

Figure~\ref{fig-16} shows how the severity of underestimating opacity depends
on both $SNR$ and resolution by comparing the mean opacity measured for the
undegraded spectra to the mean opacity measured relative to the fitted
continuum for the degraded spectra.  In Figure~\ref{fig-11}, we plot the 
distribution of the difference of these two quantities, $\Delta \tau$.
At the level of the highest
quality echelle data (top-right panel), 80-90\% of the absorption opacity is
recovered in almost every case.  At the level of poor quality HST data
(lower-left panel), the opacity underestimation is typically 50\%.
The tail to high $\Delta \tau$ in each panel represents
the relatively rare cases of heavily absorbed spectra, where no automated
procedure can adequately recover the absorption.

\section{ABSORPTION LINE MEASUREMENT}

\subsection{Line Measurement and Deblending}

We have examined the general effects of a wide range of 
resolution and $SNR$ on the appearance of the spectra and on the ability
to recover the true continuum level. For the phase of absorption line
measurement, we home in on a narrower range of parameter space with the 
four combinations: $SNR = 10$ and 30, and resolution \gamres$ = 80$ and
300 \kms. It can be seen from Figure 1 that these values bracket most
existing HST data and much of the anticipated data from the Sloan
quasar survey. This range also encompasses the data we will use in Paper
II for a Lyman-$\alpha$ coherence measurement at $z=2$.
Despite the conceptual limitations of the line/continuum fitting paradigm,
we acknowledge the fact that a huge amount of observational data has
been published using this type of procedure.  A major goal of this paper
is to form a bridge between the appropriate techniques used to analyze
simulations and the traditional methods of quasar spectroscopy.

\subsubsection{The Absorption Line Profile}

Available quasar pair data on the \Lya\ ``forest'' have $SNR$ and resolution
generally no better than $SNR = 30$ and \gamres $=80$ \kms. This corresponds 
to the highest $SNR$ and resolution of the four parameter spaces for which we 
are measuring absorption lines in the extracted spectra.  
Dectection of the weakest lines is limited by the $SNR$ and the resolution
of the data,
and measurement of the column density depends on the Doppler parameter.
However, integration of 
the evolution function for \Lya\ absorbers shows that less than 1\% of 
absorbers 
are expected to have column densities higher than $\log N=15$ \citep{sco00}.
The range of Doppler parameters for \Lya\ absorbers at $z \sim 2$ spans
approximately 20-80 \kms\ \citep{hu95}. 
For an absorber with a median Doppler value $b=30$ \kms, $\log N=15$, and
\gamres $=80$ \kms, the line profile is just saturated or reaches zero flux
at the line center.
For higher resolution data, line-fitting methodologies represent individual 
absorbers with a Voigt profile, but we demonstrate that for column densities 
less than
$\log N=15$ and resolutions lower than \gamres $=80$ \kms, the use of a 
Gaussian profile is justified.

We generate the flux profile for a \Lya\ absorption line (with essentially 
infinite $SNR$ and resolution) using subroutines from the program AutoVP 
\citep{dav97} for the case of $\log N = 15$ and $b=80$ \kms.
The convolution of this ``intrinsic'' line profile with the instrumental
line spread function (a Gaussian with \gamres) is the expected flux line 
profile. Sampling of the dispersion is chosen 
to mimic that of the degraded simulation spectra for this parameter space.  

One thousand absorption lines were created and Poisson noise was
added to represent $SNR=30$.  A Gaussian profile was fitted to each 
Monte Carlo simulated line.  This fitted Gaussian profile was compared to the 
Monte Carlo line
profile, yielding an average \chisq\ value of 0.011. For 50 degrees of freedom 
(the number of data points in the profile), this corresponds to a probability 
of less than $10^{-5}$ that the two profiles are different.  We also note that 
low resolution (\gamres $>80$ \kms) renders undetectable all the complexities 
in the absorber profile caused by hydrodynamic effects. The use of a Gaussian 
profile in fitting absorption features is therefore adequate for our immediate 
purposes, and for our initial science application in Paper II.

\subsubsection{Line Selection Algorithm}

Absorption line selection and measurement is performed using software
originally written by Tom Aldcroft \citep{ald93} that has since been
substantially revised.  The most recent version of the software retains the
original code structure, the graphic interface tools, and the fundamental
numerical subroutines.  New methodology for fitting continua, and new
algorithms for selecting and fitting absorption features have been implemented.
The continuum-fitting process is described in the previous section.  Line
selection and fitting is a two-step process where a preliminary list of lines
is selected, which forms a first estimate to the simultaneous fit that is 
determined in a subsequent step.  The line selection and fitting methodology 
was first described in \citet{pet98}.  Some refinements to the line fitting
algorithm have been added to generalize finding the techniques for a range 
of data quality and to eliminate manual intervention.
The range of data quality addressed in this work is $SNR=10$ to 30
and \gamres $=80$ to 300 \kms.

Selecting a combination of absorption lines profiles to fit an absorption 
region is a difficult because of the effects of instrumental resolution 
and $SNR$ on the true features in the spectrum. The second extraction in 
Figure~\ref{fig-8} shows 
an example (at $\sim 3662$ \AA) where lines that are easily resolved at 
20 \kms\ cannot be 
separated at a resolution 80 \kms. The $SNR$ is not as important a factor
in concealing the structure of an absorption feature until it becomes very
low. No procedure can recover 
information that is lost due to limited instrumental resolution, but 
Gaussian profile fitting does provide a fair comparison between the 
simulation spectra and observed spectra. Even if a Gaussian is a good
approximation to the shape of strong absorption features, we do not expect
that line-fitting can yield unique fits. The biggest danger
is expected to be over-fitting, such as when a single broad 
feature is fit by several overlapping Gaussians. However, it is also
possible to under-fit, such as when noise causes a weak feature to fall
below the significance threshold for a single unresolved line. In what
follows, we describe a robust procedure for dealing with most of the situations
presented by real quasar spectra.

One simple way of fitting a number of Gaussian profiles to a predefined
absorption region is to fit the peak of a Gaussian to the lowest flux point 
in the spectrum, subtract the profile from the spectrum, and then repeat 
the process at each local minimum. The method is considered to have converged
when the result is consistent with the flux error of the original spectrum. 
However, with this method, a region which may be better fit in a \chisq\ sense 
by two barely resolved lines, will actually be fit by one strong line in the
center, possibly with an unphysically large width, straddled by two smaller
``satellite'' lines. The sequential approach has the basic problem that lines
are defined one at a time, which constrains the choice of subsequent lines, and does 
not ensure an optimal or unique fit.  

We have devised an iterative method to fit the maximum number of lines to 
a region and distribute the lines optimally across the region. Unresolved 
lines are assumed, i.e. the absorber is well approximated by the instrument
line profile (a Gaussian).  Subsequently, all combinations of lines from this 
preliminary list are fit simultaneously to the region using a Marquardt
minimization technique, and the fit that meets a preset criteria is chosen 
as the best fit. We consider that the best fit is the deepest depression in 
the \chisq\ surface of solutions. The Marquardt minimization does 
not allow for large changes in the line center or line width; by design,
the preliminary step of finding lines and subsequent testing each combination 
of lines samples the surface of solutions well enough to find the best 
solution. This is justified in practice because the iterative procedure
rarely leads to substantial changes in the centers and widths of the
strongest features.

In the first phase of line selection, an iterative search is made for all
minima in the data array containing the flux values for each spectrum that 
has been convolved with the instrumental profile. A {\it primary} line list 
is formed from minima found in the first pass and a {\it secondary} list 
is made from minima found in the convolved array after lines in the {\it
primary} list are subtracted from the original data. The {\it primary} 
and {\it secondary} lists combine to form the {\it preliminary} line 
list that defines an initial ``best guess'' at the final line list. 
The rationale for combining primary and secondary lists is as follows.
The minima in the data array locate the strongest absorption features,
but these inflection points do not account for all the absorption in a
blended region. The secondary list serves to ``interleave'' the primary
list, and combination of the two yields a fairly complete and robust
map of the absorbers.

The iterative method for initially locating and measuring lines in the
spectrum of each component follows that described in detail\setcounter{n}{2}
by Paper \Roman{n} of the HST Quasar Absorption Line Key Project \citep{kp2}.
The equivalent width, $W,$ at every pixel in each spectrum is computed by
centering the spectral line profile on the $i^{th}$ pixel and performing
the weighted sum over the 6$\sigma$ limits of the Gaussian forming the array
$W_{i}$. Similarly, this is done for the error array, $\sigma W_{i}$, and 
the interpolated error array, $\bar{\sigma}W_{i}$. The interpolated error 
array is calculated by replacing flux errors for data points that deviate
negatively by more than 2$\sigma$ by the average of the errors in up to five 
(on each side) of the adjacent continuum points. Continuum in this sense 
is defined by points lying within a 2$\sigma$ deviation from the fitted
continuum. This array differs from the error array in that the errors at 
the centers of the absorption lines are increased slightly to reflect the 
noise in the adjacent true continuum. The $SNR$ of an unresolved line is
conventionally defined as $ W_{i} / \sigma W_{i} $, whereas we choose to 
define ``significance'' as $ W_{i} / \bar{\sigma}W_{i} $, which is used 
to select preliminary lines in the first phase. This correction for the 
drop in the errors at the centers of absorption lines, where the flux 
errors are smaller, allows for a more uniform comparison of relative 
line strengths.

The primary line list is generated by identifying the minima in the
$W_{i}$ array within a window of width one half the instrumental resolution
that has a significance ($W_{i} / \bar{\sigma}W_{i} $) of three or more. 
A parabolic fit through the two adjacent points identifies the line center 
and the value of $W_{i}$ and $\sigma W_{i}$ corresponding to this wavelength
give the equivalent width and error. The profiles of the lines are calculated
and subtracted from the original spectra. This subtracted spectrum is then
convolved and a secondary line list was derived in the same way. The 
secondary list is then subtracted from the original spectrum and a pass
was made again that refits only the primary lines and may add new lines
or drop existing lines. The number of lines added by the secondary list 
increases by a factor of 1.5 to 2 times 
the number of lines found in the primary list.  

It was found empirically that after approximately five iterations of 
finding lines for first the primary and then the secondary lists, the 
number of lines converged to a repeating series as the line centers 
shifted about slightly depending on the noise. The number of lines that 
are added or lost by all subsequent iterations is no more than a few 
percent of the total list. The preliminary line list was formed from 
the combination of the two lists, and if their profiles are subtracted 
from the original data array, the result has a mean flux of zero and
variations entirely consistent with noise.

\subsubsection{Line Fitting Algorithm}

The primary list of lines is input as a first guess to the simultaneous best
fit of each absorbed region in a spectrum. For practical reasons the spectrum
is broken up into regions. There are several ways of doing this and we chose
to step through each spectrum and define the region to be fit as beginning 
where the flux array downward-crosses the continuum and ending where it
upward-crosses the continuum. This method of defining the spectral regions is
reasonable for data of resolution and $SNR$ under consideration here, because 
the number of upward- and downward-crossings divides the spectrum into 
sections that can be fit by a moderate number of preliminary lines, and because
it limits the number of regions where no preliminary lines are found. Fitting 
all combinations of the preliminary lines makes it computationally impractical
to fit more than 11 lines per region, so in the rare occasions that this
situation occurs the region is split at the highest flux data point in the
region, and the fit is performed for the new region with a smaller number 
of lines.  

Lines from the initial list that fall within the region to be fit were counted,
$n$, and the number of all possible combinations of these lines, $N_{comb}$, 
was computed by summing the binomial coefficient, $C{n\choose x}$, which 
gives the number of combinations possible for $n$ lines taken $x$ positions 
at a time, over the number of possible positions, $x=1 \ldots n$
\begin{eqnarray}
 N_{comb} = \sum_{x=1}^n C {n \choose x} = \sum_{x=1}^n \frac{n!}{x!(n-x)!}.
\end{eqnarray}
The derived lines for each individual combination were worked out and saved, 
and each combination of lines was fit simultaneously to the region following 
a Marquardt minimization technique that varies the amplitude, center and 
width of the lines. The line parameters and the reduced chi square 
(\chisqnu), which is later used to select the best fit, for each combination
were saved. The method of fitting all combinations of preliminary lines
optimizes selection of the best fit to any absorption feature by maximally
utilizing the information provided by the preliminary line list.

However, in practice, \chisqnu\ is not a sensitive test of overfitting in
selecting the best fit to a region \citep{rau92}.  For example, it is possible
for a fit to produce a \chisqnu\ that is improved from the prior fit but where
the errors in the equivalent width are larger than the equivalent width itself.
In our application of this procedure, the central wavelengths are fairly well 
determined and do not change
more than a few standard deviations with subsequent fits. We therefore imposed
secondary constraints to ensure that the final fit consisted of meaningful 
line parameters. The final ``best'' fit is chosen as the fit with the lowest
\chisqnu\ that also fulfills the following criteria: 
\ (a) \chisqnu\ $\leq 100$, 
\ (b) the errors in the fitted equivalent width and the FWHM are smaller than 
the measurement itself,
\ (c) the minimum line separation for lines in a fit is $\Delta\lambda=1.12$ 
\AA, and
\ (d) 0.65\ \AA\ $<$ FWHM $<$ 4.5\ \AA. 

The first two of these constraints apply generally to the fitting procedure.
The restriction on \chisqnu\ is an empirically determined practical limitation
to the largest value expected to be associated with a reasonable fit. 
The best fit can have a \chisqnu\ that is rather high, simply due to
the fact that there may be a portion of the fit region where no lines are fit
but where the normalized flux is not exactly at the continuum level.  Failure
of this criterion forces a redetermination of the fitted region by dividing it
at the wavelength having the next highest intensity, and a refitting of the
preliminary lines to obtain a fit having a lower \chisqnu. 
The restriction on the errors of 
fitted parameters to lines prevents over-fitting.

The last two constraints depend on the range of $SNR$ and resolution of the data
being considered.  In this case, the choices were justified as follows.
The minimum and maximum FWHM values
are determined by convolving the range of expected Doppler widths \citep{hu95}
with the instrumental profile, and allowing for variation in the minimum
FWHM due to noise.  
The minimum line separation also depends on resolution and noise, and was chosen
empirically by forming the distribution of line separations and then examining
the reliability of individual fits. Instrumental
resolution in this test case corresponds to 80 \kms. There is a natural 
break in the distribution at $\sim 70$ \kms\ and lines selected with smaller
separations than this represent an over-fit with respect the features
in the undegraded simulation spectrum. Lines with separations of more than 
$\sim 90$ \kms\ are reliably selected by the algorithm. For lines with separations
between 70 and 90 \kms\, it is not clear by eye whether adding the second
line to the fit is justified, and in all cases the undegraded simulation 
spectra shows an asymmetry or two features close together. A Monte
Carlo simulation to determine the recovery rate of two close lines shows 
that the software does not reliably recover the input lines until they are
separated by $\sim$~100 \kms. This motivates our minimum separation of
1.12 \AA. However, we note that the statistical results of the analysis are
not particularly sensitive to this exact choice.

If the procedure fails one or more of the tests in any particular spectral
region, the region is split at the next highest minimum intensity point and 
a new set of simultaneous fits is made using the smaller number of lines.  
If all of these fits fail, the last fit is chosen but is tagged. The tagged
lines were examined for a subset of 200 spectra and were only 3\% of 
the total number of lines selected. Two thirds of these were moderately
broad, weak
features whose shape had been distorted with the addition of noise --- in
general, a line was found in the vicinity of the true feature even if the
addition of noise and resolution made it more difficult to detect. Almost 
all of the lines from ``failed'' fits are weak enough that they will not 
meet the selection criteria to be included in the subsequent analysis. About 
one in 5 were strong broad features where the chosen fit appeared appropriate, 
but the \chisqnu\ was too high. A remaining tiny number, barely 0.01\% 
of the whole sample, are weak features fit to the edge of broader absorption
regions or are not real features at all. This same visual inspection of many
of the fits showed that under-fitting or over-fitting is not a significant
problem for the algorithms. We conclude that the fraction of strong regions 
of absorption that are not well fit by the line/continuum algorithm is negligible. 

\begin{figure}
\epsscale{.80}
\plotone{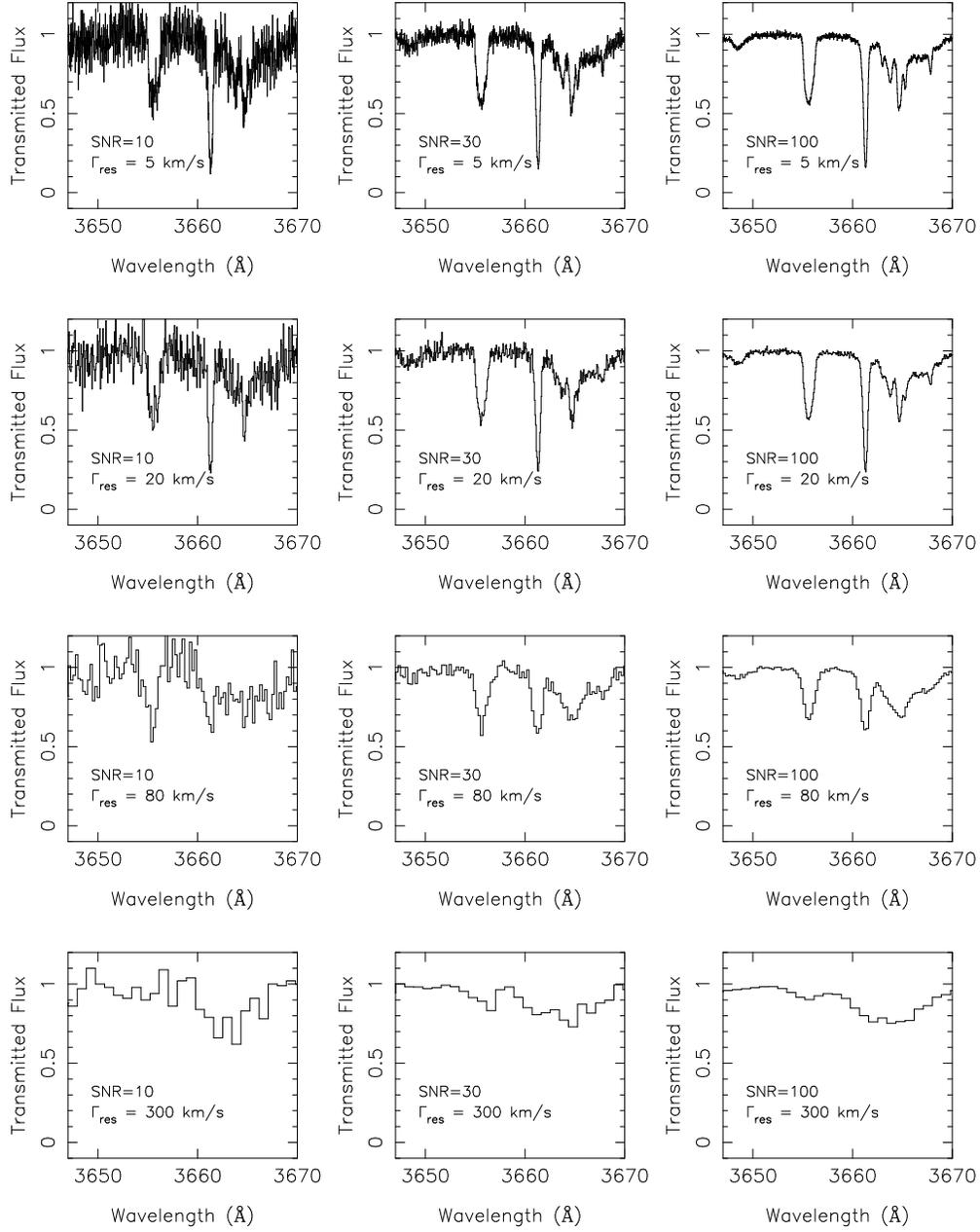}
\caption{Two spectra extracted from the simulation, each of typical 
opacity at $z=2$. The twelve panels in each case show the spectrum 
degraded to $SNR=10,30, \rm and\ 100$, and \gamres $=5,20,80, \rm and\ 300$ 
\kms. The top right panel is a reasonable facsimile of the original
spectrum.
\label{fig-8}}
\end{figure}

The procedures for line-fitting are robust and can be applied in an
automated way, but they are also somewhat complex.  Figure~\ref{fig-12}
shows how they work in practice, as applied to the two extracted spectra 
from Figure~\ref{fig-8}, and in each of the four combinations of $SNR = 10$
and 30, \gamres $=80$ and 300 \kms.  In other words, Figure \ref{fig-12} shows
actual line/continuum fits to the four parameter choices at the lower left
of Figure~\ref{fig-8}.  Fitted continua and lines are overplotted in each 
case with the location of significant ($5\sigma_{det}$) absorbers shown by tick marks.
``Significance'' is defined in two ways:  $\sigma_{det}$ is related to the
flux error and is used to describe the strength of a line in terms of the 
detection limit of the data; 
$\sigma_{fit}$ is the error in the equivalent width returned by the software
and is a measure of line reliability or goodness of fit.  In practice,  
$\sigma_{det}$ can be
used to set a uniform detection threshold for absorbers and is directly related to 
the $SNR$, where $\sigma_{fit}$ is an indicator of how well the equivalent width
is determined
--- lines with large $\sigma_{fit}$ might be found in noisy regions of the 
spectrum or in the wings of broader absorption features.  All lines fit by
the software are retained regardless of strength or reliability, but for the
analysis we define two categories: {\it secure} and {\it marginal}.
The criterion for inclusion in either list is $5\sigma_{det}$. Lines that are
$< 5\sigma_{fit}$ make up the {\it marginal} list, and lines that have an equivalent
width $\ge 5\sigma_{fit}$
constitute the {\it secure} list.  Marginal lines may be reliable enough to be
useful in the comparison with observed spectra, but in this work we only
consider lines from the {\it secure} list in the analysis.
In Figure \ref{fig-12}, longer, thicker tick marks show the location of the secure lines,
and shorter, narrower tick marks denote lines from the marginal category.

\section{PROPERTIES OF THE ABSORBERS}

\subsection{Line Counts Along Single Sight-lines}

Absorption features were measured using the line-fitting algorithm for 300
primary lines of sight (PLOS) and for their adjacent sight-lines spanning
separations from 33.3 to 1000 $h_{50}^{-1}$ kpc  --- a total of 2100 spectra
extracted from the simulation.
This was done for two different signal-to-noise ratios, $SNR = 
10$ and 30, and two resolutions \gamres $= 80$ and 300 \kms, forming 
four realizations of the data. 
We compared the distribution of absorber properties measured from the 
simulated spectra to those found from observations of large samples 
of spectra published in the literature as a basic cross-check of our method.

In Figure~\ref{fig-14}, we show the distribution of the line centers of absorbers 
in 300 PLOS of the $SNR=30$, \gamres$=80$ \kms\ realization for each of the three
projection axes.  The deviations from the mean (shown as the dotted line) indicate
large scale, coherent structures within the simulation box, having filamentary
and sheetlike properties.  However, the average of the 3 projections converges
toward the mean value indicating that the absorbers approach a uniform distribution
over the full size of simulation box, and demonstrating that there are no artifacts 
introduced by edge effects.
Additionally, we examine the distribution of the number of absorbers per line
of sight in Figure~\ref{fig-15}.   This distribution is consistent with a 
Gaussian distribution that is the expectation for a random distribution of absorbers.
In particular, the high multiplicity of absorbers that would indicate substantial
clustering is not seen.  

Table~\ref{tbl-1} lists the average number density ($dN/dz$) of secure lines per line 
of sight above a limiting rest equivalent width threshold ($5\sigma_{det}$) for 
each of the four realizations of the simulated spectra (column 5). 
For Sample 1 (column 1) we used 1000 PLOS for increased statistics on the highest
$SNR$ and resolution realization, and for the other three samples we used 300 PLOS.  
The number of lines from observations is 
obtained in two ways from a study performed by \citet{sco00}.  These values are 
listed in columns 6 and 7 of Table~\ref{tbl-1}. First, we simply count the number 
of absorbers found in the small wavelength range spanned by the simulated spectra 
at $z=2$ (J. Scott, private communication). However, a limited number of absorbers 
are found in this 23 \AA\ wide region and so the error bars are large. Since
the evolution function (described below) is close to linear over $2<z<2.1$, the 
Poisson error can be reduced by using the lines counted in this five times larger 
spectral region as an estimate for the smaller spectral region actually spanned
by the simulation. 

Second, the average number of lines at $z=2$ can be computed from coefficients 
of an evolution function fitted to an appropriate sample of observed absorbers. 
This method is less direct but it has the merit of incorporating most of the
available data.  The number of \Lya\ lines per unit redshift per unit equivalent 
width can be described as follows: 
\begin{eqnarray}
\frac{\delta^2 \cal{N}}{\delta z \delta W} = \frac{A_0}{W^*} (1+z)^{\gamma}
\exp \left(-\frac{W}{W^*}\right).
\end{eqnarray}
Integrating Equation 2 with respect to $W$ gives the number of lines per line
of sight with a rest equivalent width greater than 0.16 \AA\ for the chosen values
${\cal{A}}_0 = 5.86$, $W^*=0.257$, and $\gamma=2.42$ from \citet{sco00}:
\begin{eqnarray}
\frac{\delta \cal{N}}{\delta z} = {\cal{A}}_0 (1+z)^{\gamma},\
\rm where \  {\cal{A}}_0=A_0 \exp\left(-\frac{W}{W^*}\right).
\end{eqnarray}
Further integration with respect to $z$ gives the number of lines 
per spectrum in the range $z_{min} < z < z_{max}$:
\begin{eqnarray}
N = \frac{{\cal{A}}_0}{\gamma+1} \left[ (1+z_{max})^{(\gamma+1)}
-(1+z_{min})^{(\gamma+1)} \right].
\end{eqnarray}
The number of lines in this redshift interval for a different equivalent
width limit, $w_{lim}$, is obtained by scaling $N$ by the factor 
\begin{math}\exp \left[ - \frac{(w_{lim}-0.16)}{W^*} \right].
\end{math}
Errors on $N$ are obtained using the $1\sigma$ error bars on $\gamma$.
See column 7 in Table~\ref{tbl-1}.

A comparison is made only for the \gamres $=80$ \kms\ realizations (Samples
1 and 3), because this resolution closely matches the data set used by 
\citet{sco00}.  The error bars listed in Table~\ref{tbl-1} are merely reflect
the Poisson error and do not include systematic uncertainties due to differences
in the line-fitting algorithms that can be as large as 10-20\%.  
Sample 3 most closely matches the sensitivity of the data set used by \cite{sco00},
and the number of lines found in the extracted spectra agree to within 20\%
with both methods for counting observed lines.
Sample 1 in Table 1 agrees with the (poorly-determined) number predicted using
a model for the evolution of the absorbers, and it also agrees to within 20\%
of the expected number from the direct line counting method.

In addition to a check of aggregate line counts, it is important to check whether
extracted spectra from the simulation return the observed distribution of line
strength.  We therefore compare the number of lines per equivalent width bin, 
$dN/dW$, to a measurement that uses the data of 
\citet{sco00}. This recent study provides better statistics at $z=2$ 
than the Hubble Space Telescope Absorption Line Key Project \citep{kp1,wey98}.
Figures~\ref{fig-13} and \ref{fig-2} are plots of the number of secure 
($\ge 5\sigma_{fit}$) absorbers counted 
in the spectra from the simulations, binned by rest equivalent width, for Samples 1 and 3. 
Both distributions are well within the $1\sigma$ error bars 
on $\gamma$, showing that the absorber statistics obtained by line-fitting
applied to the simulation spectra are consistent with those from observed
samples. Distributions formed that include the marginal lines ($<5\sigma_{fit}$) 
show that the number of lines continues to increase in accordance with the 
exponential distribution down to $2.5\sigma_{fit}$, and inspection of 
Figure~\ref{fig-12} shows that these {\it marginal} lines 
always correspond to a real feature in the ``true'' spectrum.
We conclude that absorber line counting from this hydrodynamic simulation provides an
excellent match to the demographics of the Lyman-$\alpha$ forest at $z=2$.

\subsection{Coincident Lines Between Sight-lines}

The first indications that Lyman-$\alpha$ absorbers were larger than the
halos of individual galaxies came from the detection of coincident lines
in the spectra of quasar pairs \citep{din94, bec94}. Most paired lines of 
sight experiments define absorbers pairs by choosing a maximum separation 
in velocity, typically 50-300 \kms, that lines must have in order to be 
labeled as coincident and therefore physically associated. The operational
definition of a matching ``window'' depends
on the line density, and it is chosen to minimize the probability of a 
chance match. A ``coincident'' pair is two lines that match in velocity closely
enough that the probability of a chance match is small.  However, the preselection 
of a velocity match window can
potentially lose information, and it makes an implicit assumption about
the kinematic state of the absorbing gas (i.e. the amplitude of any velocity
shear on that particular transverse scale). A very large number of paired
lines of sight can be extracted from the simulations, with many potential
absorber pairs at each of the transverse separations. We have designed a
matching algorithm that does not depend on any prior definition of a velocity
match.  This allows an examination of how the coherence of the absorbers 
changes over transverse separations of 33 \hfifty\ kpc to 1 \hfifty\ Mpc.

Coincident lines are defined such that coincidences are symmetric. In other
words, if there is a line in sightline $A$, its coincident line is the 
nearest line in velocity space in sightline $B$. However, coincident lines
are also selected starting with sightline $B$ and matching to sightline $A$.
An absorber is not identified as a coincident line unless it is a reversible
match both from $A$ to $B$ and from $B$ to $A$. This procedure will of course
result in lines that are unmatched in both sight-lines. We account 
for the fact that the simulation cube is periodic by matching across the
ends of each simulation spectrum. To discuss line matching statistics, we 
have chosen to use the single realization of the simulation spectra with 
$SNR$ = 30 and \gamres$=80$ \kms. This most closely mimics the observational
spectra of the three quasar pairs at $z=2$ 
that we will analyze in Paper II. Absorber matches were found for 300 primary 
sight-lines --- seven extracted spectra make up each set.   The fraction of
matched lines is 87\%, 73\%, 65\%, 61\%, 59\% and 58\% for transverse
separations 33, 100, 233, 400, 667, 1000 \hfifty\ kpc, respectively.
The fraction of matched lines for random lines of sight formed by pairing
lines of sight from the x-projection with lines of sight from the z-projection
is 57\%.

Figure~\ref{fig-4} illustrates the line-matching procedure. Figure~\ref{fig-4}$a$
is merely a test of the algorithm.  Figure~\ref{fig-4}$b$ shows the effect of noise 
and subsequently slight differences in the degraded spectra on the scatter 
between the equivalent widths of matched lines. At a $SNR=30$, only 3\%
of the lines fail to match due to fluctuations added by random noise. 
Figure~\ref{fig-4}$c$ shows that
there can be substantial differences in line strength even across
a transverse scale of 33 \hfifty\ kpc, which is below the resolution of the simulation
where intrinsic differences must be small.  Nevertheless, the formal correlation
is highly significant. 
Figure ~\ref{fig-4}$d$ shows the null experiment, where absorbers detected 
in 300 primary sight-lines are matched against 300 randomly selected 
sight-lines  from an orthogonal projection of the simulation.  This ensures
that any two features at similar wavelengths are separated by 10-20
\hfifty\ Mpc in space, and so are expected to be uncorrelated.

In Figure~\ref{fig-4}$b$, the scatter is larger than anticipated due to the 
addition of noise because the line selection procedure must deal with issues of 
deblending and continuum fitting. The very few outliers (out of 924 line
pairs) are mostly caused by the situation where the addition of noise to a
marginally resolved feature causes it to be fit with two components in one
sightline but only one in the other. At first sight, the large scatter in
Figure~\ref{fig-4}$c$ is more surprising, because the transverse separation is 
at the limit of resolution of the simulation. The scatter comes about because 
the mapping from real space to spectral space in complex. Peculiar gas motions
affect the optical depth both by shifting line centers and by changing line
profiles due to velocity gradients (Bi \& Davidsen 1997). The fact that our
line fitting software is responding to real physical differences in opacity
can also be seen from the analytic modelling of Viel et al. (2001). They
show calculated \Lya spectra at $z = 2.15$ with noticeable differences 
on transverse scales as small as 60 \hfifty\ kpc, where a significant 
fraction of lines fit with Voigt profiles have column densities that 
differ by more than a factor of two. Taken together with the added
effects of noise, the scatter in Figure~\ref{fig-4}$c$ can be readily 
understood.

A major result of this paper is shown in Figure~\ref{fig-5}.  This shows
matched or coincident lines plotted against the velocity difference of the
line pair for the six transverse separations with respect to the PLOS.
(Of course, it is possible to generate many transverse separations between
0 and 1000 \hfifty\ kpc using these spacings, but the choices in 
Figure~\ref{fig-5} illustrate the coherence phenomenon adequately.)
Given the mean line spacing of 850 \kms\ (for a $5\sigma$ line), 
the line matching algorithm becomes aliased at 425 \kms, so matches with 
separations above this value are not physically meaningful.

The most sensitive test of absorber coherence uses the fact
that truly coincident lines will have a small velocity separation between
sight-lines, where random matches (not physical associations) will have a
larger range of velocity differences. Figure~\ref{fig-5} suppresses the
information on homogeneity by plotting the average of the equivalent 
widths of the paired absorbers against the velocity splitting of the pair.  
The distribution of the velocity splittings of matched pairs changes quite
dramatically with transverse separation. To quantify this, we plot the
cumulative distribution of the velocity splittings for the paired absorbers 
in Figure~\ref{fig-6}. To show the expectation for a random set of absorbers
with no physical association between the sight-lines, we also form the
cumulative distribution for pairings between spectra extracted from
orthogonal projections through the simulation volume. Any two paired 
absorbers in this case will have a physical separation of $\sim$10-20 
\hfifty\ Mpc and so are not anticipated to show coherence (the dashed lines 
in Figure~\ref{fig-6}). A K-S test demonstrates that there is an excess 
of small velocity splittings, and therefore detectable coherence, up to 
about 500 \hfifty\ kpc.

A typical method of examining the homogeneity of Lyman-$\alpha$ absorbers 
over various transverse separations is to plot the equivalent widths of 
each line in a matched pair. However, noise and intrinsic differences in 
the absorbing regions imprint themselves on a spectrum and affect the line 
fitting algorithm in a complex way, so that apparently similar spectra can 
yield coincident line pairs with substantially 
different equivalent widths. 

To see if line matching depends on line strength, we formed distributions
of equivalent widths for both the paired lines as well as the unpaired 
lines. A K-S test shows these distributions are drawn from the same
distribution. Even though there are many more weak (lower column 
density) lines than strong (higher column density) lines, the degree
of coherence on any particular transverse scale does not depend on line
strength. However, the fraction of matched lines decreases with increasing
transverse scale; a clear demonstration that coherence is being lost as
the scale approaches 1 \hfifty\ Mpc.

\subsection{Comparison with Continuous Statistics}

Up until now we have been examining the coherence of the \Lya\ absorbers
using the highest density peaks that are fitted as lines.
Clustering or coherence may also be analyzed through the use of continuous
statistics that leverage the flux information in every independent point of the
spectra.  Various types of correlation measures are well described in \cite{cen98}.
We define the two-point correlation function (TPCF) of the normalized absorbed
flux, $F(v)$, along neighboring lines of sight as
\begin{eqnarray}
\xi_{flux}(v)=\frac{\left<F_1(v+v')F_2(v')\right>}{\left<F\right>}-1,
\end{eqnarray}
where $F_1$ and $F_2$ refer to fluxes along the two lines of sight.

Figure~\ref{fig-17} shows the TPCF for each of the transverse separations
33, 100, 233, 400, 667, and 1000 \hfifty\ kpc.  We compute the auto-correlation
function, which is the PLOS correlated with itself ($F_1=F_2$ in Equation 5), 
and this is shown as the
curve with the highest amplitude.  We also compute the TPCF for lines of
sight that are randomly associated by correlating the PLOS from spectra
extracted from the x-projection with PLOS from the z-projection.

The TPCF for discrete absorbers in a line-counting experiment is computed for 
all transverse separations by comparing 
the number of observed nearest neighbor pairs, $N_{obs}$, with the number of pairs 
formed from random lines of sight, $N_{ran}$ as follows: 
\begin{eqnarray}
\xi_{pair}=\frac{N_{obs}}{N_{ran}} -1.
\end{eqnarray}
The random lines of sight are the PLOS from the x- and z- projections.
Figure~\ref{fig-17} shows the correlation functions for each of the six
transverse separations.  The amplitude of the TPCF for each method 
decreases to approach the curve for random demonstrating that coherence
can be measured using line-fitting methods.
In Paper II, we will illustrate the different types of information that are
gleaned by using continuous flux statistics and line-counting techniques,
and we will quantify the amount of data required to detect coherence using
line counting.

\section{SUMMARY}

This paper has established techniques for the intercomparison of quasar spectra and  
spectra extracted from cosmological simulations, recognizing that observations are  
traditionally interpreted in terms of line-counting, while simulations offer a direct  
measure of neutral hydrogen opacity at every resolution element. This work anticipates  
a series of direct comparisons using space- and  ground-based observations of quasar  
pairs and groups. Initial tests have been carried out on the \Lya\ forest at   
$z = 2$, with simulation spectra degraded to span the range of observational
quality of typical ground and space-based data.
The main results are as follows:

(1) Simulators measure opacity with respect to a pre-defined continuum, while observers
must determine a quasar continuum that is not known {\it a priori}, in the presence of 
noise and emission features. Software has been designed that robustly measures continua  
both for the small wavelength (redshift) range of simulations and for the long wavelength
range of typical observations. The systematic underestimate of opacity ranges from  
$\Delta\tau = 0.02$ for $SNR = 100$, \gamres$= 5$ \kms\ to $\Delta\tau = 0.06$ for 
$SNR = 10$, \gamres$ = 300$ \kms.

(2) A fully automated procedure has been developed for the detection and selection of
absorption features that produces reliable results for 
observational data having $SNR=10$ to 30 and \gamres $=80$ to 300 \kms.
The techniques are robust even in extended regions of
strong absorption. A direct comparison with published quasar surveys shows that the  
number density and equivalent width distribution of absorption lines at $z = 2$ agrees 
with measurements of spectra extracted from an SPH simulation.

(3) A technique has been established to match absorption lines between adjacent lines
of sight. The match rate of coincident lines is used to search for coherence in the 
absorbing gas on transverse scales up to 1 \hfifty\ Mpc. The two-point correlation
function of matched pairs reveals a significant excess 
out to transverse scales of $\sim$ 500 \hfifty\ kpc, indicating the detection of  
\Lya\ coherence on this scale. Most of the signal of excess pairs occurs with
velocity splittings of $< 100$ \kms, indicating that the velocity field is quiet. 
The coherence signal measured with line counting and matching techniques agrees well 
with results from a two-point flux correlation analysis, which uses all the information 
in the simulation spectra.

We are particularly grateful to Tom Aldcroft, who made available the code that
formed the core of the software suite described in this paper, and Jennifer
Scott, who shared her thesis data. We acknowledge useful discussions with 
Romeel Dav\'{e}, Craig Foltz, Jane Charlton, and Rupert Croft. 
This work was supported
by NASA Astrophysical Theory Grants NAG5-3922, NAG5-3820, and
NAG5-3111, by NASA Long-Term Space Astrophysics Grant NAG5-3525, and
by the NSF under grants ASC93-18185, AST-9803072, and AST-9802568.

\acknowledgments




\begin{deluxetable}{crrrrrr}
\footnotesize
\tablecaption{Comparison of Number Density of \Lya\ Absorbers \label{tbl-1}}
\tablewidth{0pt}
\tablehead{
\colhead{Sample} & \colhead{$SNR$}   & \colhead{\gamres\tablenotemark{a}}   &
\colhead{$W_{lim}$\tablenotemark{b}} &
\colhead{${\left(\frac{dN}{dz}\right)}_{sim}$} &
\colhead{${\left(\frac{dN}{dz}\right)}_{obs}$\tablenotemark{c}}  &
\colhead{${\left(\frac{dN}{dz}\right)}_{obs}$\tablenotemark{d}} \\
\colhead{} & \colhead{}   & \colhead{(\kms)}   &
\colhead{(\AA)} &
\colhead{}  &
\colhead{}  &
\colhead{} \\
\colhead{(1)} & 
\colhead{(2)} & 
\colhead{(3)} &
\colhead{(4)} &
\colhead{(5)} &
\colhead{(6)} &
\colhead{(7)} \\
}
\startdata
1 & 30 & 80   & 0.05  &  $170 \pm 3 $ & $ 143 \pm 11 $ & $128^{+125}_{-~63}$   \\
2 & 30 & 300  & 0.05  &  $ 75 \pm 4 $ & \nodata & \nodata  \\
3 & 10 & 80   & 0.17  &  $ 92 \pm 4 $ & $ 108 \pm 2 $ & $81^{+79}_{-40}$   \\
4 & 10 & 300  & 0.17  &  $ 38 \pm 3 $ & \nodata & \nodata   \\
\enddata

\tablenotetext{a}{Resolution of the degraded simulation spectra as described by the 
FWHM of a Gaussian distribution in \kms.}  
\tablenotetext{b}{The limiting rest equivalent width for an isolated 
$5\sigma_{det}$ absorber.  
}
\tablenotetext{c}{The number of absorbers per unit redshift obtained by
counting the number of lines in the sample of \citet{sco00} in the 
redshift range $2 < z < 2.1$ with $5\sigma_{fit}$.  The errors are $1\sigma$ Poisson 
errors; systematic uncertainties are likely to be larger.}
\tablenotetext{d}{The 
number of absorbers per unit redshift computed from the absorber evolution 
model of \citet{sco00}.  The errors correspond to the $1\sigma$ errors on $\gamma$,
where the integrated evolution function is specified by Equation 4 in the text.}

\end{deluxetable}



\clearpage




%


\begin{figure}
\epsscale{.80}
\plotone{f8d.eps}
\end{figure}


\begin{figure}
\epsscale{.80}
\plotone{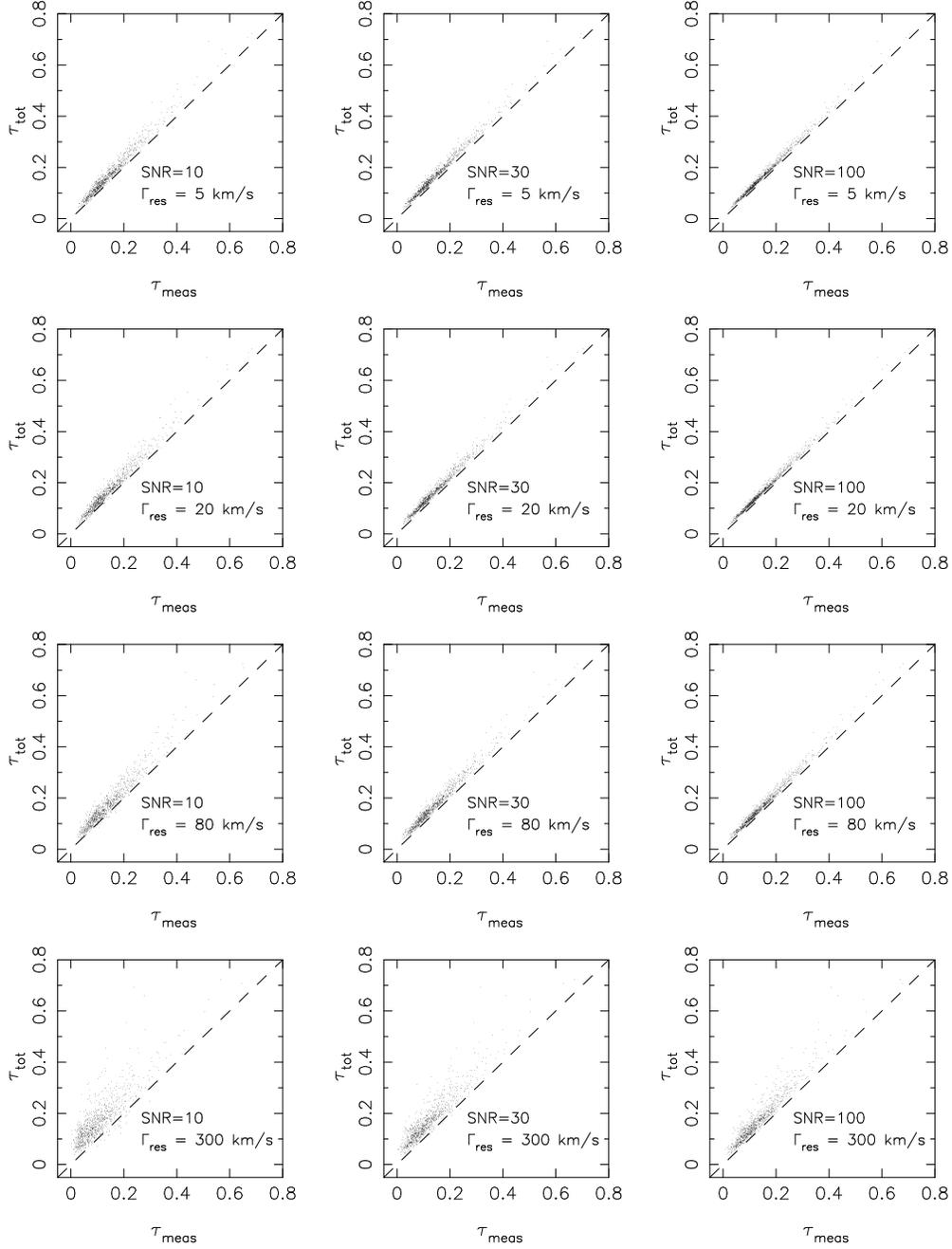}
\caption{An illustration of how the underestimate of opacity due to 
continuum fitting depends on both $SNR$ and resolution.  The mean opacity
from each undegraded spectrum is plotted against the mean opacity measured
relative to the fitted continuum for each degraded spectrum.
Each panel represents a different combination of
$SNR$ and resolution; the layout is the same as Figure~\ref{fig-8}.
\label{fig-16}}
\end{figure}

\begin{figure}
\epsscale{.75}
\plotone{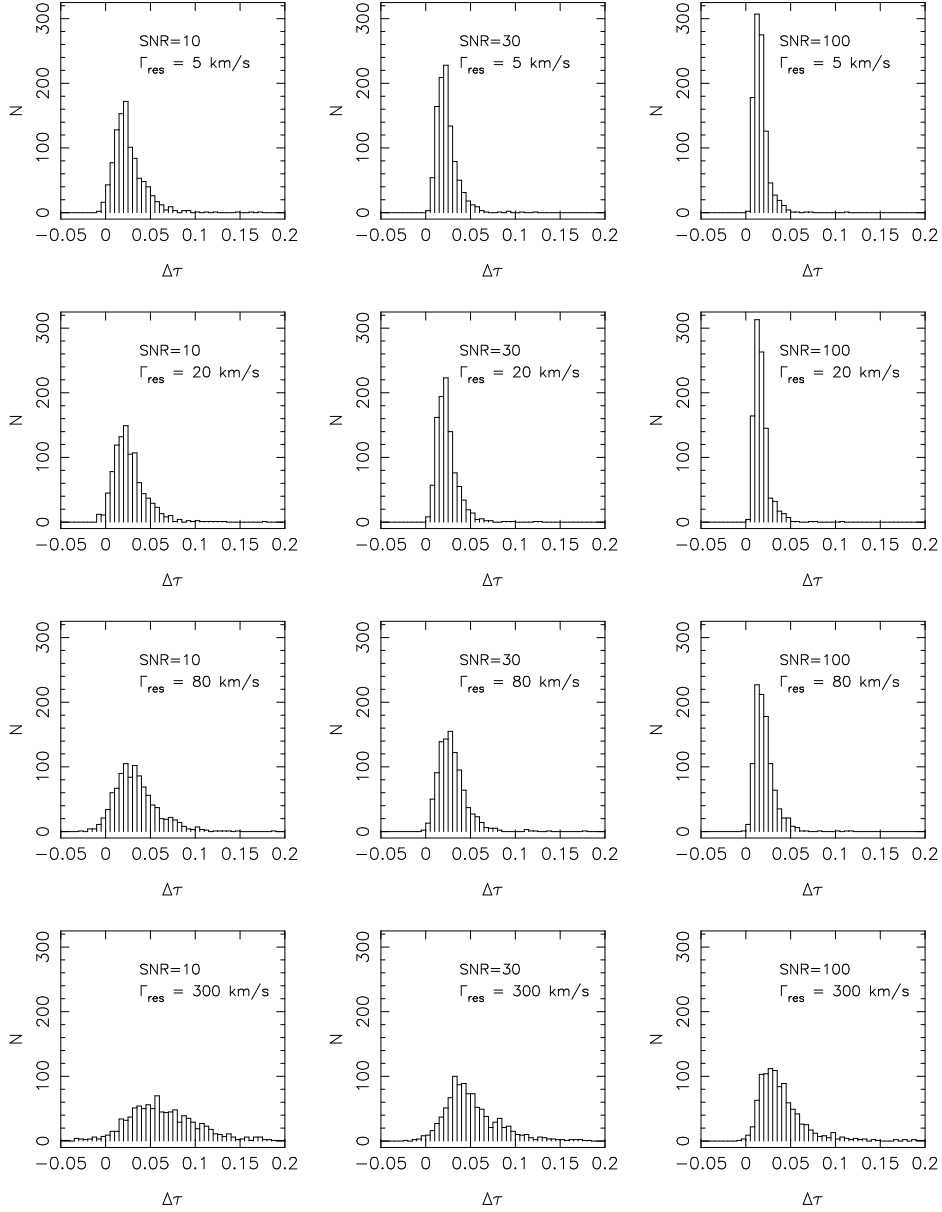}
\caption{
The opacity decrement from the continuum fitting procedure applied to 1000
spectra extracted from the simulation.  The quantity $\Delta\tau$ is the 
difference between the continuum level of an input spectrum (raw, undegraded) 
and the fitted continuum to a degraded spectrum, 
converted into opacity. 
\label{fig-11}}
\end{figure}

\begin{figure}
\epsscale{.75}
\plotone{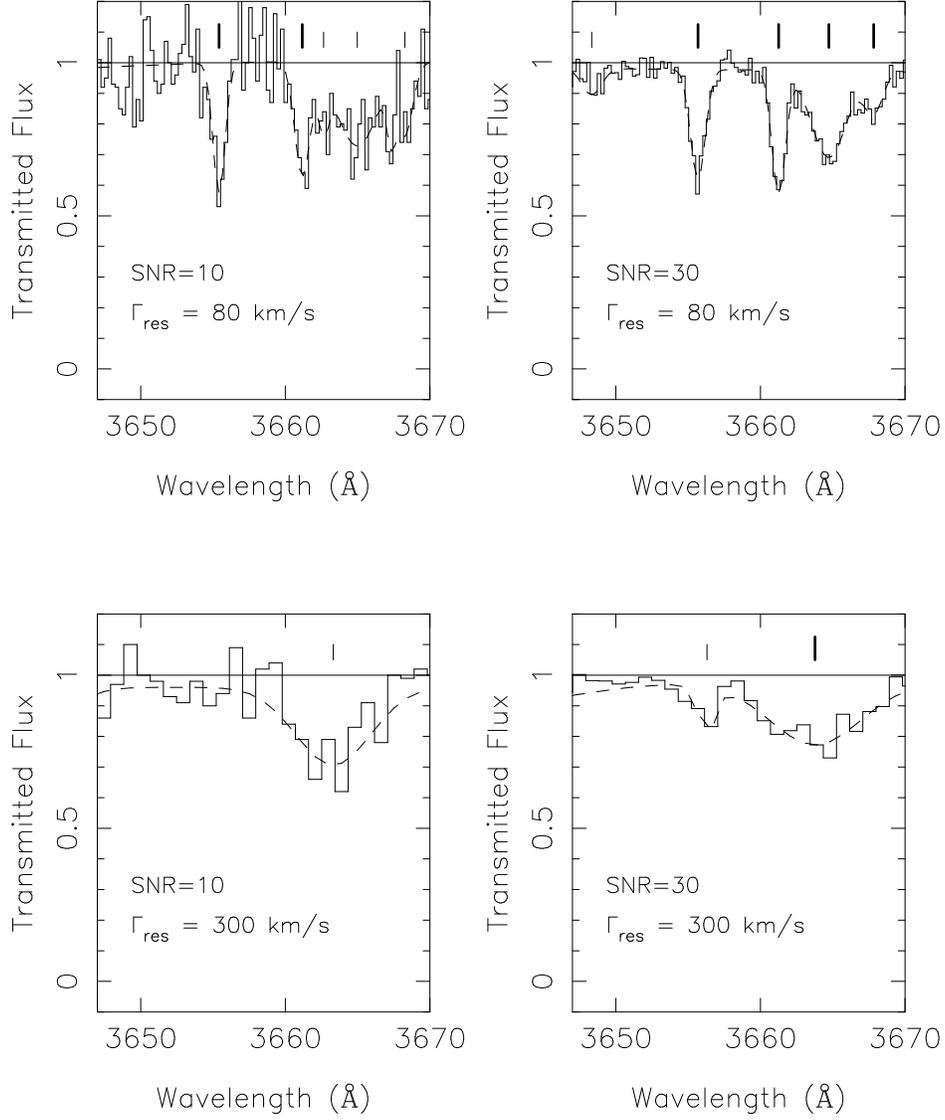}
\caption{
An illustration of the line-fitting procedure for the two extracted spectra 
from Figure~\ref{fig-8}, in each of the realizations $SNR=10\ \rm and\ 30$, 
\gamres$=80$ and 300 \kms.  Fitted continuum and superimposed lines are shown 
by a dashed line, and line centers of absorbers with strength of at least
$5\sigma_{det}$ are shown by tick marks.  Longer thicker ticks denote lines 
with secure reliability $\ge 5\sigma_{fit}$ and shorter thinner ticks mark 
lines with marginal reliability of 
$< \sigma_{fit}$.  \label{fig-12}}
\end{figure}
\begin{figure}
\epsscale{.75}
\plotone{f12d.eps}
\end{figure}

\begin{figure}
\epsscale{.75}
\plotone{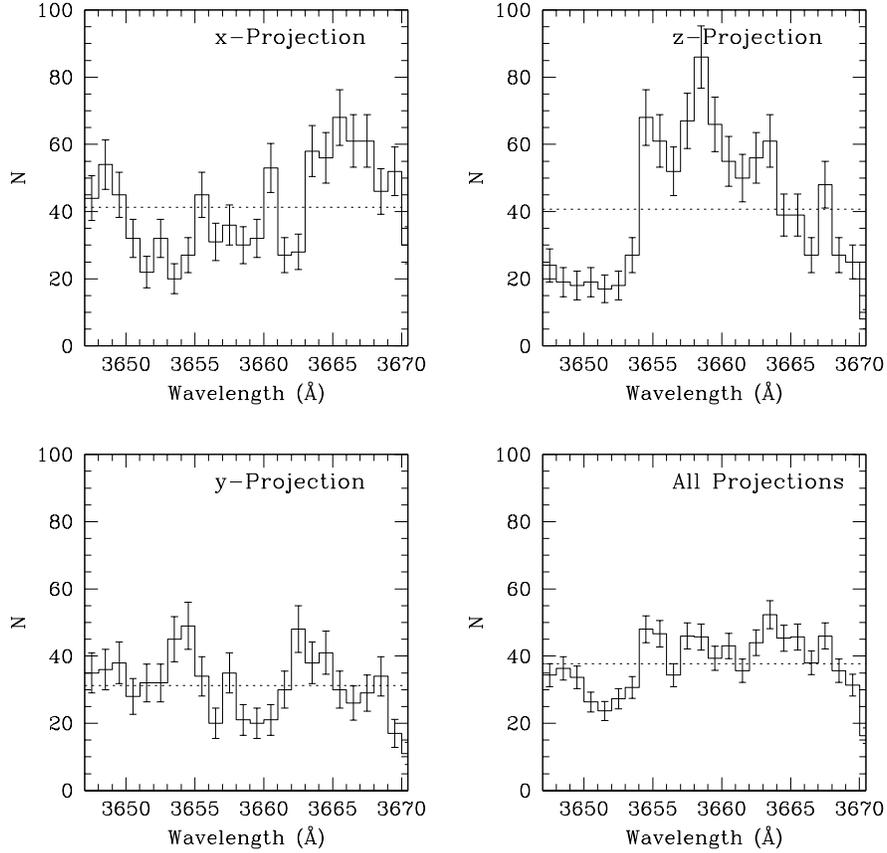}
\caption{The distribution of line centers for absorbers found in 300 PLOS for
the $SNR=30$, \gamres$=80$ \kms\ realization for the x-, y-, and z- projections.
The dotted line shows the mean of the distribution.  The individual projections
reveal coherent, large-scale structures.  The average of the individual projections 
({\it lower right}) 
shows that the distribution converges toward a uniform distribution as more lines of 
sight from all projections are combined.
\label{fig-14}}
\end{figure}

\begin{figure}
\epsscale{.75}
\plotone{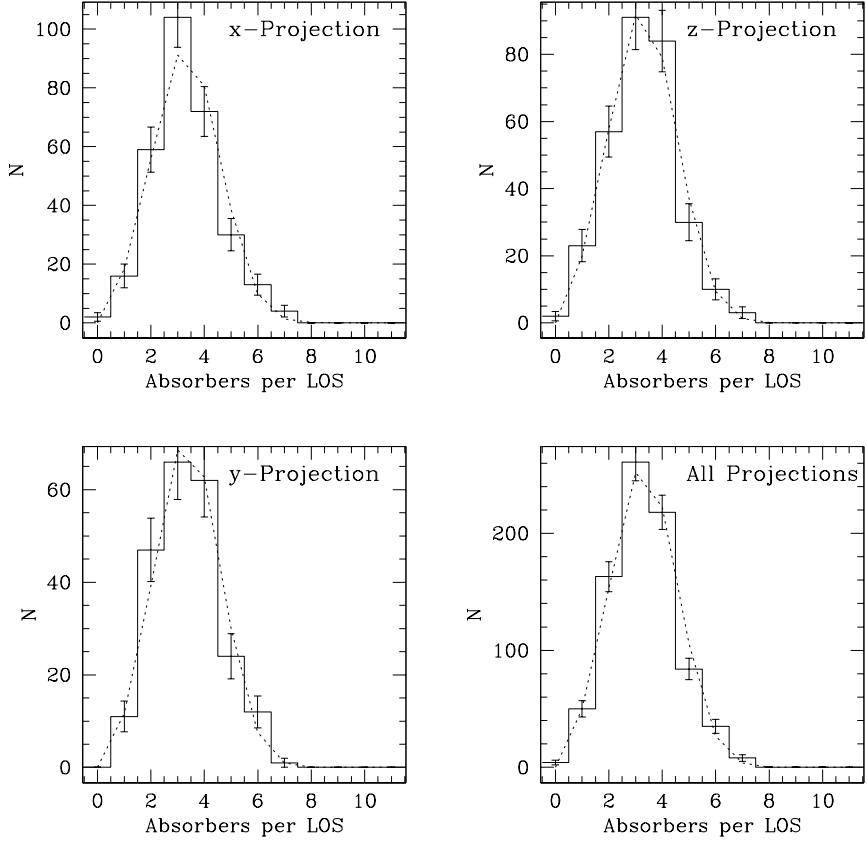}
\caption{The distribution of the number of absorbers per line of sight for 
300 PLOS for the $SNR=30$, \gamres$=80$ \kms\ realization for the x-, y-, and z- 
projections. The dotted line is a Gaussian distribution with the mean and 
standard deviation of the data.  The good match to the Gaussian demonstrates
that the number of absorbers per line of sight is consistent with a random distribution
of absorbers on scales similar to the size of the simulation box.
\label{fig-15}}
\end{figure}

\begin{figure}
\epsscale{.65}
\plotone{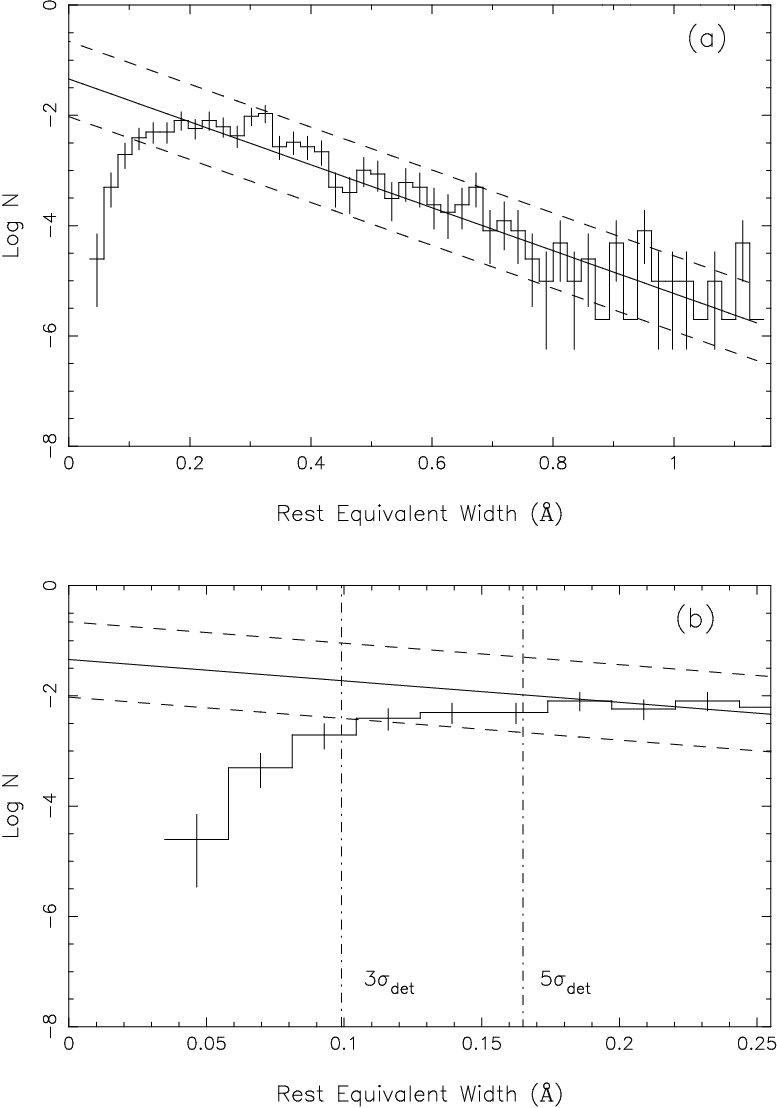}
\caption{$(a)$ A histogram of the number of $5\sigma_{fit}$ absorbers as 
a function of rest
equivalent width, measured by line-fitting applied to 300 spectra extracted
from the simulation degraded to $SNR=10$, \gamres$=80$ \kms. 
The solid line is the best-fit evolution function to
the observed data at $z=2$ from \citet{sco00}, with dashed lines showing $1\sigma$ error 
bars due to the parameter $\gamma$.  $(b)$ Expanded view of the distribution of weak line
strengths, with detection significance superimposed.
\label{fig-13}}
\end{figure}

\begin{figure}
\epsscale{.65}
\plotone{f2.eps}
\caption{$(a)$ A histogram of the number of $5\sigma_{fit}$ absorbers as 
a function of rest
equivalent width, measured by line-fitting applied to 1000 spectra extracted
from the simulation degraded to $SNR=30$, \gamres$=80$ \kms. As in Figure~\ref{fig-13},
the solid line is the best-fit evolution function to
observed data at $z=2$ from \citet{sco00}, with dashed lines showing $1\sigma$ error 
bars due to the parameter $\gamma$.  $(b)$ Expanded view of the distribution of weak line
strengths, with detection significance superimposed.
\label{fig-2}}
\end{figure}

\begin{figure}
\epsscale{.80}
\plotone{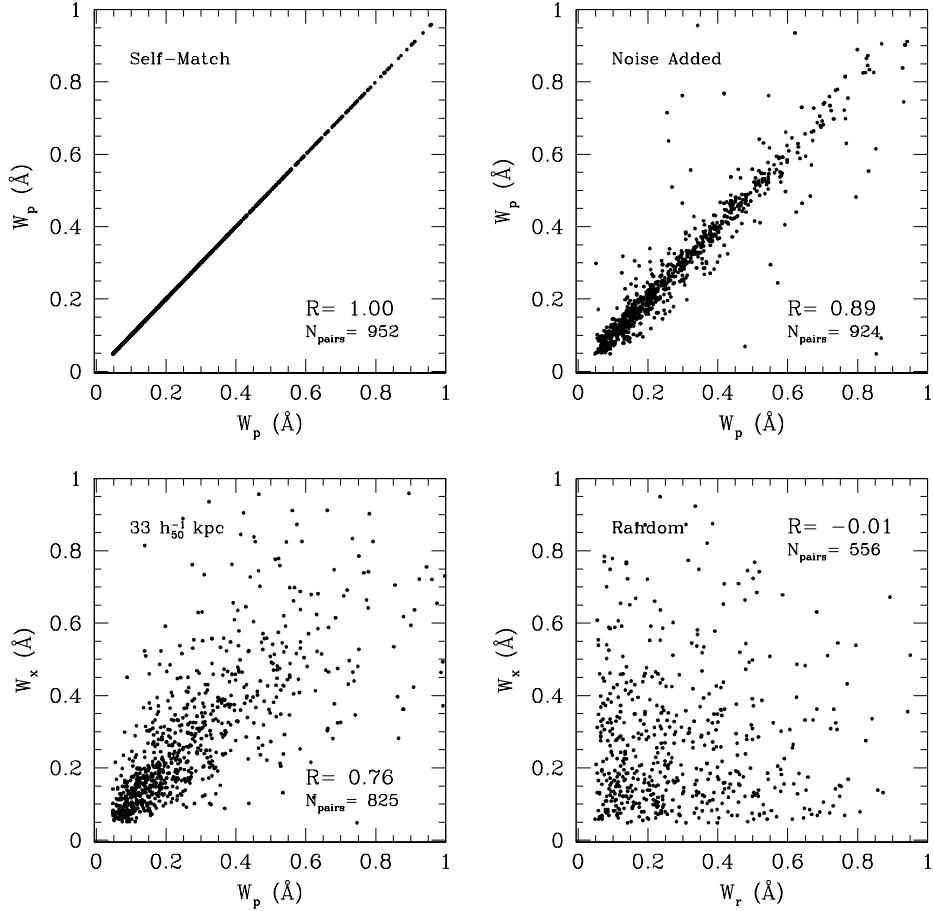}
\caption{A plot of the rest equivalent widths of paired absorbers for the
sample where $SNR=30$ and \gamres $=80$ \kms. $(a)$ 300 primary
lines of sight are matched to themselves as a test of the algorithm, so the 
correlation is perfect.  $(b)$ 
300 primary lines of sight are matched to themselves, but with a
different noise seed.  $(c)$ Matched pairs found for the closest transverse
separation of 33 \hfifty\ kpc, which approaches the resolution of the simulation, so
intrinsic differences in the spectra should be small. $(d)$ 300 primary lines
of sight are matched to 300 randomly-selected lines of sight from an orthogonal
projection of the simulation.  No pairs due to physical coherence
are expected in this situation.
\label{fig-4}}
\end{figure}

\clearpage

\begin{figure}
\epsscale{.80}
\plotone{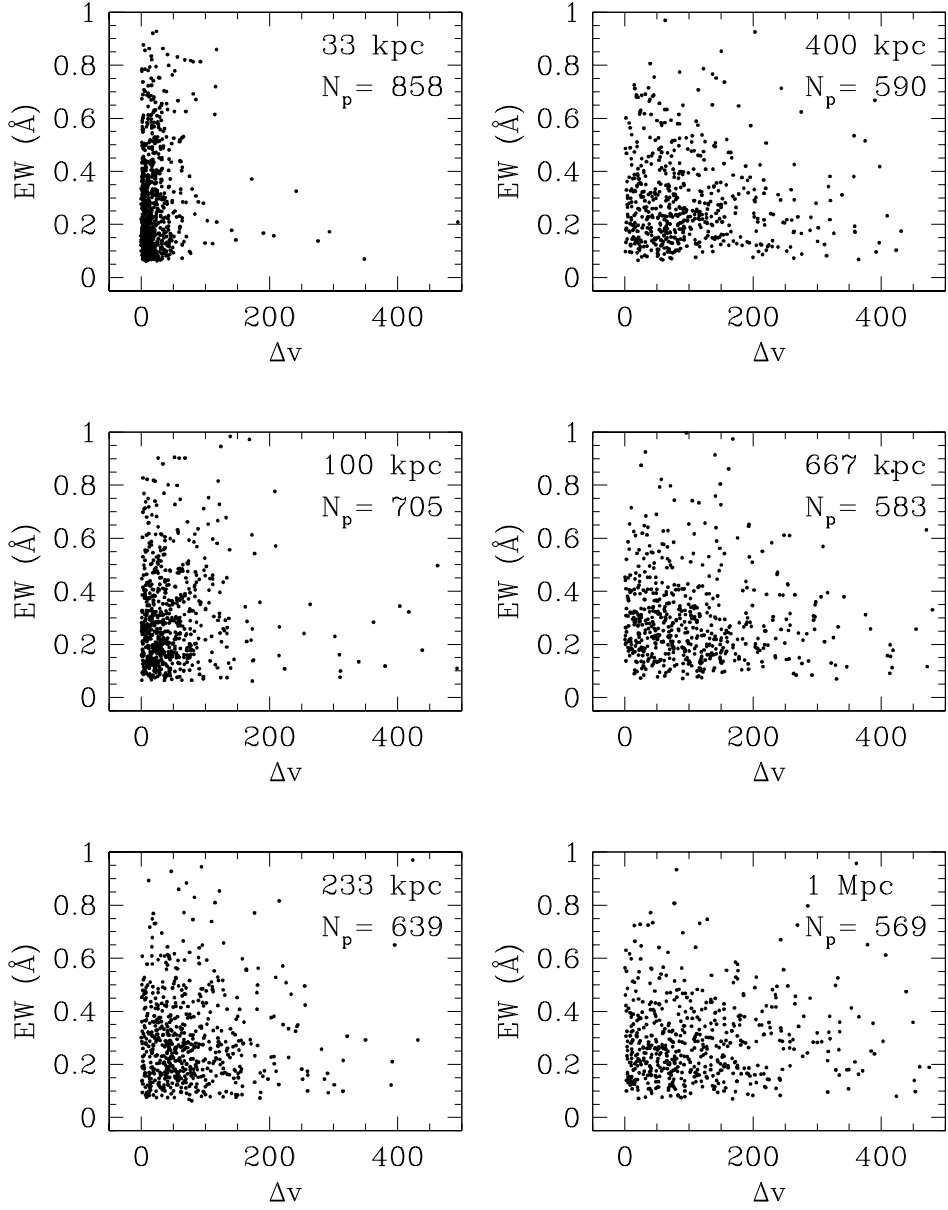}
\caption{The average rest equivalent width for matched absorber pairs plotted
against the velocity splitting of the pair for 300 sets of lines of sight at
6 transverse separations.  The fraction of matched lines goes down from
87\% for a separation of 33 \hfifty\ kpc to 58\% for a separation of 1000
\hfifty\ kpc.  The fraction of matched lines for random lines of sight is 57\%.
\label{fig-5}}
\end{figure}

\clearpage

\begin{figure}
\epsscale{.80}
\plotone{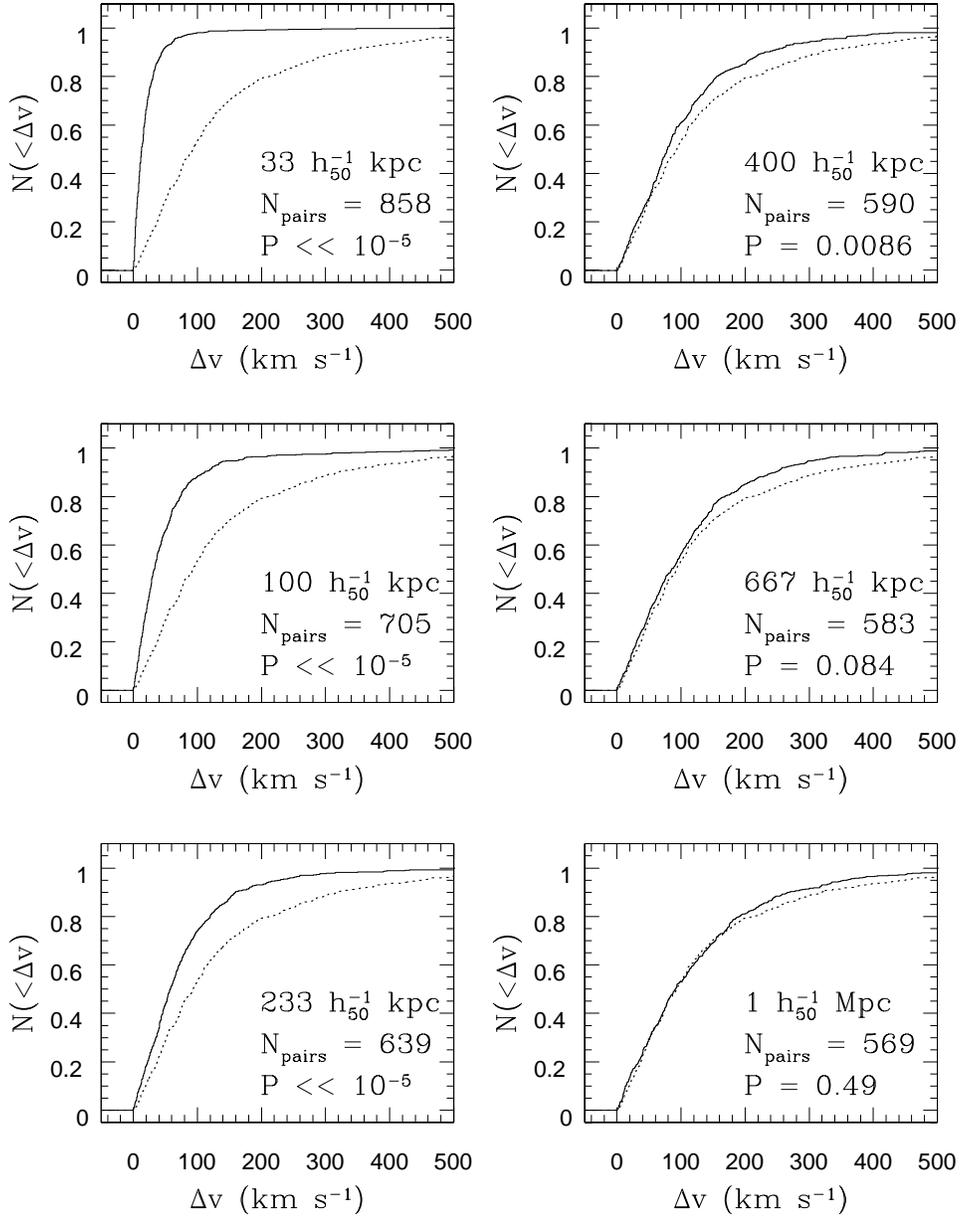}
\caption{The cumulative distribution of velocity splittings for absorber pairs
at each transverse separation in the sample (300 sets) is shown by the solid
line.  The dashed line is the cumulative distribution of absorber pairs formed
by pairing x-projection primary lines of sight with z-projection primary lines
of sight, approximating the distribution expected for random absorber pairs.
\label{fig-6}}
\end{figure}

\clearpage

\begin{figure}
\epsscale{.70}
\plotone{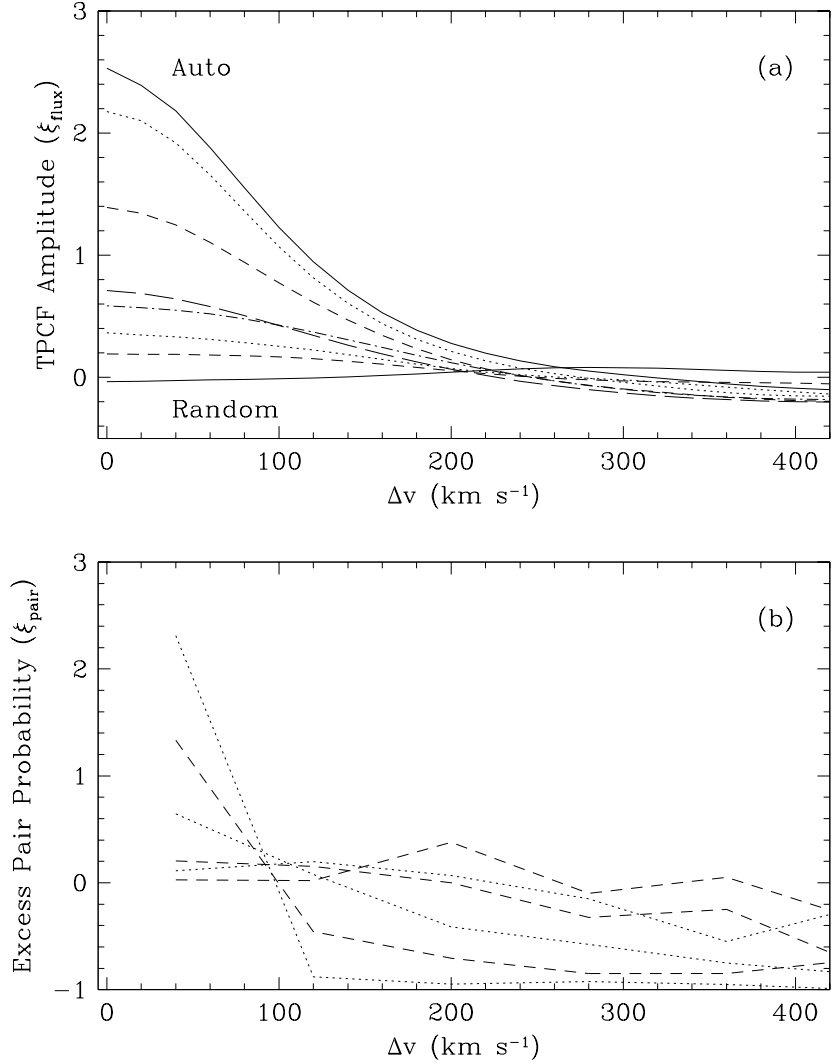}
\caption{{\it (a)} The two-point correlation function (TPCF) of the transmitted
flux for spectra having transverse separations of 33, 100, 233, 400, 667 and 1000 
\hfifty\ kpc, which is the sequence from the highest to the lowest curve at $\Delta v=0$.  
The curve with the highest amplitude is the auto-correlation formed from correlating 
the PLOS with themselves.  The TPCF for random lines of sight is formed by correlating the 
PLOS of the x-projection with the PLOS of the z-projection.
{\it (b)} The TPCF for discrete absorber pairs.  The curve with the highest
amplitude is for a transverse separation of 33 \hfifty\ kpc.  Curves with subsequently 
lower amplitudes are for separations of 100, 233, 400, 667 and 1000 \hfifty\ kpc. 
\label{fig-17}}
\end{figure}

\end{document}